\documentclass[aps,pra,superscriptaddress,amsmath,amssymb,amsfonts,twocolumn,nofootinbib,longbibliography]{revtex4-2}

\usepackage{amssymb}
\usepackage{dcolumn}
\usepackage{graphicx}
\usepackage{bbm}
\usepackage{amssymb,amsmath}
\usepackage{amsmath}
\usepackage{amsfonts}
\usepackage{hyperref}
\usepackage{color}
\usepackage{soul,xcolor}
\usepackage[normalem]{ulem}
\usepackage{bm}
\usepackage{braket}
\usepackage{amsmath}
\usepackage{graphicx,color,xcolor,colortbl}
\newcommand{\be}{\begin{equation}}
\newcommand{\ee}{\end{equation}}
\newcommand{\beq}{\begin{eqnarray}}
\newcommand{\eeq}{\end{eqnarray}}

\newcommand{\hide}[1]{}

\begin{document}

\title{Tunable pairing with local spin-dependent Rydberg molecule potentials in an\\ atomic Fermi superfluid }

\author{Chih-Chun Chien}
\email{cchien5@ucmerced.edu}
\affiliation{Department of Physics, University of California, Merced, CA, USA}

\author{Seth T.~Rittenhouse}
\affiliation{Department of Physics, the United States Naval Academy, Annapolis, Maryland 21402, USA}
\affiliation{ITAMP, Center for Astrophysics $|$ Harvard $\&$ Smithsonian, Cambridge, USA}

\author{S.~I.~Mistakidis}
\affiliation{ITAMP, Center for Astrophysics $|$ Harvard $\&$ Smithsonian, Cambridge, USA}
\affiliation{Department of Physics, Missouri University of Science and Technology, Rolla, MO, USA}

\author{H.~R.~Sadeghpour}
\affiliation{ITAMP, Center for Astrophysics $|$ Harvard $\&$ Smithsonian, Cambridge, USA}

\begin{abstract}
We explore the energy spectrum and eigenstates of two-component atomic Fermi superfluids 
with tunable pairing interactions in the presence of spin-dependent ultra long-range Rydberg molecule (ULRM) potentials, within the Bogoliubov-de Gennes formalism. The attractive ULRM potentials lead to local density accumulation, while their difference results in a local polarization potential and induces the in-gap Yu-Shiba-Rusinov (YSR) states whose energies lie below the bulk energy gap. 
A transition from equal-population to population-imbalanced occurs as the pairing strength falls below a critical value, accompanied by the emergence of local Fulde-Ferrell-Larkin-Ovchinnikov (FFLO) like states characterized by out-of-phase wave functions and lower energies compared to the YSR states. The negative contribution emanating from the FFLO-like states also causes a sign change in the gap function within the ULRM potentials. Depending on the Rydberg excitation, the transition towards population-imbalance can be on either the BCS or the Bose-Einstein condensation side of the Fermi superfluid. Additionally, spin-polarized bound states arise along with oscillatory ``clumpy states" to compensate for the local density difference. 
We discuss possible experimental realization  of the composite Rydberg atom-Fermi superfluid system.
\end{abstract}
\maketitle

\section{Introduction}
The conventional BCS theory has singlet pairing which resists spin population imbalance~\cite{Tinkham_book,Fetter_book}. Nevertheless, there have been continual studies on BCS superfluids against spin-polarization potentials. For a homogeneous two-component Fermi gas with attractive interactions, the Chandrasekhar-Clogston limit~\cite{10.1063/1.1777362,PhysRevLett.9.266} states that the BCS superfluid phase is no longer stable when an external uniform magnetic field exceeds a critical value. Although the original Chandrasekhar-Clogston limit considers the scenario that Zeeman splitting drives the BCS superfluid to a spin-polarized normal Fermi gas, later the possibility of superfluids with spatially modulated order parameters, known as the Fulde-Ferrell-Larkin-Ovchinnikov (FFLO) states~\cite{PhysRev.135.A550,LO64}, have gained significant attention. 
Moreover, the FFLO state is supposed to survive in the ground state of 1D attractive Fermi gases under a uniform polarization potential~\cite{PhysRevResearch.3.043105}. 
Evidences for the existence of FFLO states range from condensed matter~\cite{STEGLICH1996498,Imajo2022,doi:10.1126/science.abb0332,Wan2023} to cold-atom~\cite{Liao2010,PhysRevLett.117.235301} settings, see also Refs.~\cite{PhysRevB.51.9074,doi:10.1143/JPSJ.76.051005,PhysRevB.73.214527,PhysRevB.75.014521} for corresponding theoretical investigations.

In addition to the studies on BCS superfluids subjected to  uniform polarization potentials, it is known that localized, in-gap Yu-Shiba-Rusinov (YSR) states may arise in a BCS superfluid under the influence of a spatially localized magnetic field or a spin-polarization potential~\cite{YU1965,Shiba68,Rusinov69}. 
As argued in Ref.~\cite{YU1965}, the origin of YSR states is traced back to the spin-exchange term of the polarization potential. This is due to the fact that the expectation value of the average potential $(V_{\uparrow}+V_{\downarrow})(n_\uparrow+n_{\downarrow})/2$ with respect to the BCS pair wave function depends on whether the potentials are attractive or repulsive, but that of the polarization potential $(V_{\uparrow}-V_{\downarrow})(n_\uparrow-n_{\downarrow})/2$ is negative. Here, $V_{\sigma}$ and $n_{\sigma}$ refer to the local spin-dependent potentials and densities of the $\sigma=\uparrow,\downarrow$ states. 
Therefore, the energy levels of the Cooper pairs are negatively shifted by the polarization potential. When the few lowest-energy states fall below the energy gap, they form the in-gap YSR states featuring  localized wave functions. 
The original works regarded YSR states induced by repulsive spin-dependent potentials, while later works suggested applications in superconductor~\cite{RevModPhys.78.373,HEINRICH20181} and cold-atom~\cite{PhysRevLett.128.175301,PhysRevA.105.043320} systems. Moreover, spin-dependent contact attractive potentials have been argued to induce in-gap YSR states in BCS superfluids~\cite{PhysRevA.83.033619} as well. 

Here, we propose a versatile platform to interrogate the response of Fermi superfluids experiencing local spin polarization and address the above phenomena locally by analyzing Rydberg atom-Fermi superfluid composite systems with tunable interactions and potentials. The concept is based on
ultra long-range Rydberg molecule (ULRM) formation due to the scattering between the Rydberg electron and an adjacent ground-state atom with suitable conditions~\cite{PhysRevLett.85.2458,Chibisov02,Hamilton02}. 
ULRMs have been experimentally realized ~\cite{Bendkowsky2009,Niederprum2016,Althon2023,Booth15,PhysRevA.101.060701,Shaffer2018,Fey20} upon immersing bosonic Rydberg atoms in a background of bosonic atoms.   In addition, there have been theories on the formation of Bose and Fermi polarons exploring 
their quantum bunching and anti-bunching behavior associated with Rydberg molecules~\cite{PhysRevLett.120.083401,PhysRevLett.116.105302,PhysRevA.97.022707,Sous20,Durst24}. Meanwhile, theoretical predictions of ULRMs from fermionic Rydberg atoms with noninteracting fermions~\cite{Sous20} and Fermi superfluids across the BCS-Bose Einstein condensation (BEC) crossover~\cite{RM_PRA24} unveil the intriguing behavior 
of Rydberg molecules due to the Fermi statistics and Cooper pairs. 

While the ULRM potentials of the composite Rydberg atom-Fermi superfluid systems used in Ref.~\cite{RM_PRA24} were assumed to be spin-independent,  
spin-dependent ULRM potentials have been calculated in selected systems~\cite{PhysRevA.95.042515,PhysRevA.98.042706}. Here we obtain the spin-dependent ULRM potentials of potassium atoms within 
first order perturbation theory in the presence of s-wave electron-atom contact interactions. 
Specifically, the spin-dependent ULRM potentials introduce a local spin polarization potential acting as effective Zeeman splitting for the Fermi superfluid.

Furthermore, deploying the Bogoliubov-de Gennes (BdG) framework~\cite{degennes-sc,BdG-book} for the quasi-1D Fermi superfluid in the presence of spin-polarized ULRM potentials, we find a transition from equal-population to population-imbalance as the pairing strength decreases. The location of the transition can be on either the BCS or the BEC side, depending on the Rydberg-atom state of the spin-polarized ULRM potentials. 
Moreover, the transition is accompanied by a sign-change of the gap function within the potentials, reminiscent of the modulation of the order parameter of the population-imbalanced FFLO superfluid~\cite{PhysRev.135.A550,LO64}. 

The Rydberg atom-Fermi superfluid composite systems allows to assess special states that are different from typical bulk states.
Concretely, we will show the following types. (1) The spin-polarized YSR states with energies below the energy gap of the Fermi superfluid and localized wave functions. (2) The spin-polarized bound states localized in the ULRM potential wells and  stemming from broken Cooper pairs (due to the attractive ULRM potentials) with energies usually higher than the energy gap of the Fermi superfluid. (3) A collection of clumpy states at higher energies, sometimes above the depths of the ULRM potentials to offset the population imbalance due to the first two types of localized states. Those clumpy states with oscillatory wave functions in between localized and delocalized states mainly originate from the opposite spin-component of the YSR states. (4) The FFLO-like states being the energetically lower ones and contributing a negative gap function within the local spin-polarization potential through a phase shift between its wave functions and population imbalance of the Fermi superfluid.

The rest of the paper is organized as follows. Sec.~\ref{Sec:Potential} introduces the spin-dependent ULRM potentials of 
potassium Rydberg atom-Fermi superfluid systems. Sec.~\ref{Sec:BdG} describes the BdG formalism for obtaining the gap function and densities of the Fermi superfluid under the spin-dependent ULRM potentials. Sec.~\ref{Sec:Kn50} analyzes the case  
of the K (50S) ULRM potentials, elaborating on the  
transition from equal to imbalanced population 
on the BCS side and also presenting the energy spectrum and wave functions of emergent special states. 
Sec.~\ref{Sec:Kn40}  generalizes our results 
for the stronger K (40S) ULRM potentials. 
In Sec.~\ref{Sec:Exp}, we argue about possible experimental probes for the Rydberg atom-Fermi superfluid system. 
We offer concluding remarks and discuss future perspectives in  Sec.~\ref{Sec:Conclusion}. 

\section{Spin-dependent ULRM potentials}\label{Sec:Potential}
In the following, we consider a few Rydberg atoms immersed in a two-component Fermi superfluid and focus on the formation of Rydberg molecules by the spin-dependent ULRM potentials. 
There are different ways to realize such Rydberg atom - Fermi superfluid systems. For example, it is possible to introduce one of the  $^{39}$K or $^{41}$K isotopes into an atomic Fermi superfluid consisting of two hyperfine states of $^{40}$K atoms and excite the bosonic $^{39}$K or $^{41}$K atoms to create Rydberg states which they subsequently produce the spin-dependent ULRM potentials. 
Experimentally, Bose-Fermi mixtures of $^{39}$K-$^{40}$K~\cite{Bochenski24} and $^{41}$K-$^{40}$K~\cite{PhysRevA.84.011601} have been realized. Alternatively, 
one may solely consider $^{40}$K atoms initially prepared in three different hyperfine states. Two of these hyperfine states, e.g.  $|\uparrow\rangle=|9/2,-7/2\rangle$ and $|\downarrow\rangle=|9/2,-9/2\rangle$, experience pairing interactions and form a superfluid while the atoms residing in the third state, e.g. $|3\rangle=|9/2,7/2\rangle$, are excited into Rydberg states to generate the ULRM potentials. 
We remark, however, that if some of the $^{40}$K atoms from the Fermi superfluid (i.e., in $|\uparrow\rangle$ or $|\downarrow\rangle$ states) are excited into Rydberg states, instead of introducing additional atoms to produce the ULRM potentials, the excitations from the superfluid will suppress the pairing gap due to the pair breaking in the superfluid~\cite{RM_PRA24}.

The interaction between a Rydberg atom and its nearby fermions~\cite{Sous20} takes the form  $V_{Ryd,\sigma}(x)\psi_\sigma^{\dagger}(x)\psi_\sigma(x) d^\dagger d$, where $d$ ($d^{\dagger}$) represents the annihilation (creation) operator of a Rydberg atom, 
$V_{Ryd,\sigma}(x)$ is the ULRM potential for spin $\sigma$, and
 $x$ measures the distance from the Rydberg atom to the scattered atom. 
Around a Rydberg atom, $d^\dagger d$ may be replaced by $\langle d^\dagger d\rangle =1$, thereby rendering $V_{Ryd,\sigma}(x)$ 
an effective potential for the spin $\sigma$ atoms in the Fermi superfluid. When $V_{Ryd,\uparrow}(x)\neq V_{Ryd,\downarrow}(x)$, they act as an effective, position-dependent Zeeman splitting term.

The ULRM potential arises due to the 
interactions between a Rydberg atom and a ground-state ${}^{40}$K atom within the Rydberg orbit.  The Hamiltonian describing this process is given by
\begin{equation}
H=H_0 + V_{e-^{40}K} + 
 + H_{hf},
\label{eq:totalH}
\end{equation}
where $H_0$ is the Hamiltonian for the Rydberg atom. The hyperfine Hamiltonian is ${H}_{hf}=A_{hf} \mathbf{s}_g \cdot \mathbf{i}_g$, where $\mathbf{i}_g$ is the ground state nuclear spin and $A_{hf}$ is the hyperfine constant; for $^{40}$K, $A_{hf}$ = -285.731 MHz~\cite{touchard1982}, and $i_g=4$.
Here the interaction between the Rydberg electron and the ground state atoms  is given by the Fermi contact pseudopotential, $V_{e-^{40}K}=\sum_{S,M_S}\frac{2\pi \hbar^2 a^S(k)}{m_e}\delta(\vec{r}-\vec{x}) \ket{SM_S}\bra{SM_S}$, where $\vec{r}$ and $\vec{x}$ are the positions of the Rydberg electron and the ground state atom respectively. Note that the interaction potential depends on the total spin of the Rydberg electron and the ${}^{40}$K ground state valence electron in the electronic singlet ($S=0$) or triplet ($S=1$) states.  The effective ${}^{40}$K spin dependence of the URLM arises from the coupling with the $^{40}$K ground hyperfine term \cite{anderson2014}. 


The electron scattering interaction strength is obtained by the  energy-dependent scattering length, $a^S(k)=a^S_0+\pi \alpha k/3$, where $\alpha=300$ a.u. is the ${}^{40}$K polarizability in atomic units \cite{eiles2019trilobites}, $k=\sqrt{2m_e E/\hbar^2}$ and $E$ is the scattering energy. Here, we approximate the energy semi-classically as the electron kinetic energy in the Coulomb potential, $E=e^2/x-E_b$ where $E_b$ is the electron binding energy of the K($50S$) or K($40S$) Rydberg state and $e$ is the electron charge. The zero-energy scattering length, $a^S_0$ depends on whether the Rydberg electron is in the singlet, $S=0$, or triplet, $S=1$ state with the ${}^{40}$K valence electron.  For this work, we use $a^{S=0}_0=+0.55 a_{b}$  and $a^{S=1}_0=-15 a_{b}$ \cite{karule1965elastic}, where $a_{b}$ is the Bohr radius. 

\begin{figure}[t]
\centering
\includegraphics[width=\columnwidth]{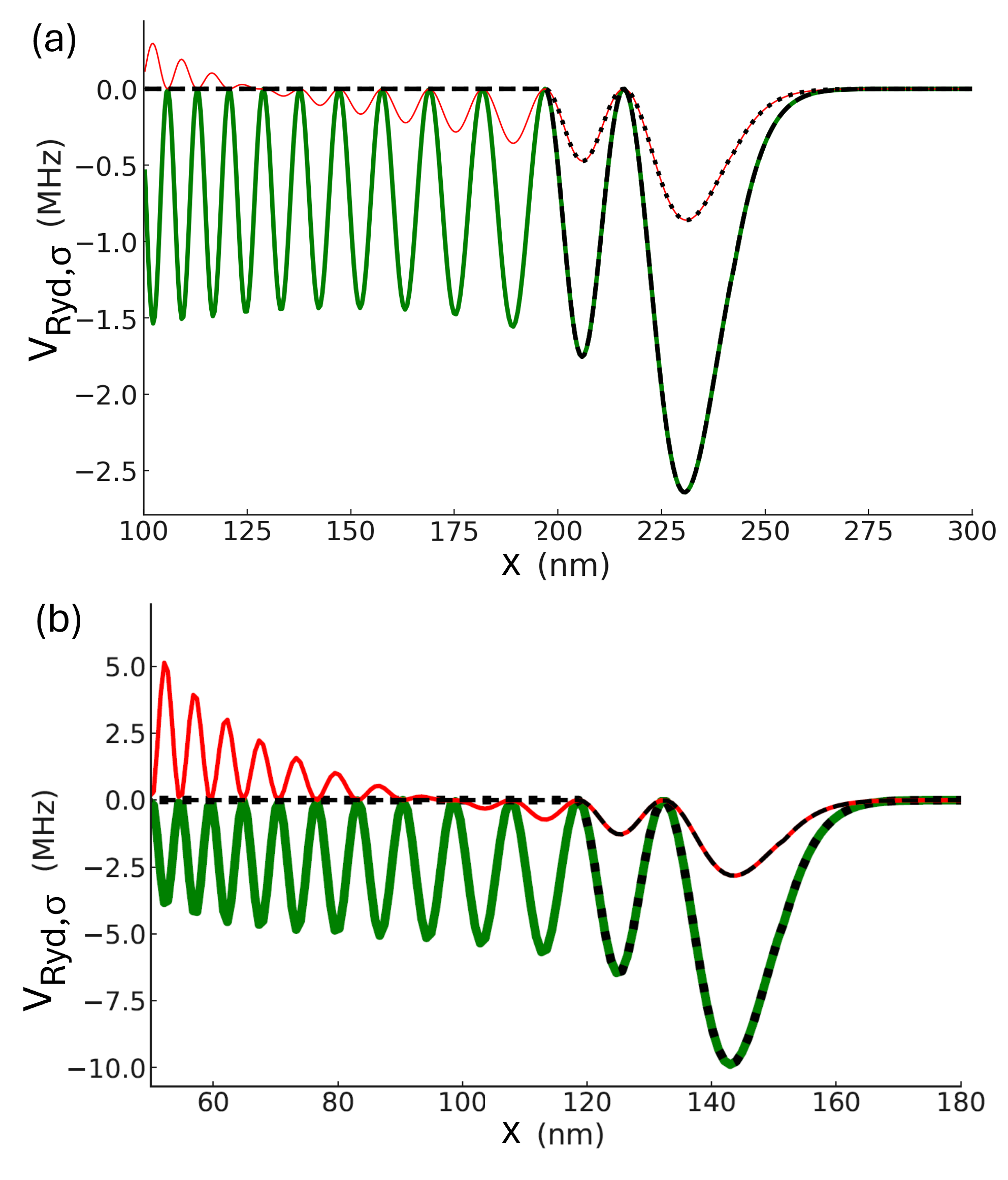}
\caption{Spin-dependent ULRM potentials of (a) K($50S$) state and (b) K($40S$) Rydberg state of a potassium atom for the two hyperfine states of a $^{40}$K atomic superfluid. The deeper (shallower) potential corresponds to $V_{Ryd,\uparrow}$ ($V_{Ryd,\downarrow}$) while the dashed and dotted lines represent the double-well approximations of the spin-dependent ULRM potentials, respectively.}
\label{Fig:K40Pot}
\end{figure}

The $nS$ Rydberg states of ${}^{40}$K are energetically isolated separated from other Rydberg states by several GHz for the $40S$ and $50S$ states considered here \cite{lorenzen1983quantum}. Thus the ULRM potentials can be obtained via degenerate perturbation theory by averaging over the spatial degrees of freedom in the electron-atom interaction assuming that the Rydberg electron is in the nS Rydberg state and the ground state atom's position is fixed.  The remaining, spin-dependent Hamiltonian is then diagonalized in the basis of states where the total angular momentum projection along the inter-nuclear axis is $M=m_s+m_f=-4$ where $m_s=\pm 1/2$ is the Rydberg electron spin and $m_f=-7/2,-9/2$ is the internuclear projection of the ${}^{40}$K in the $f=9/2$ hyperfine state. The resulting ULRM potentials $V_{Ryd,\sigma}$ are shown in Fig.~\ref{Fig:K40Pot} for the K(40S) and K(50S) Rydberg states. Note that we have subtracted the Rydberg energy and hyperfine energies so that $V_{Ryd,\sigma}(x\rightarrow\infty)=0$. In general, the Rydberg interaction potentials correspond to states of mixed hyperfine spin.
However, in the case of ${}^{40}$K, the deeper potential corresponds to a $\sim90\%$ admixture of the $\ket{\uparrow}$ state, while the weaker one consists of a $\sim 90\%$ $\ket{\downarrow}$ admixture.  For the purposes of this work, we will assume that each potential corresponds to a pure $\ket{\uparrow}$ or $\ket{\downarrow}$ state. Moreover, using $^{39}$K as the Rydberg atom in a $^{40}$K superfluid generates basically the same ULRM potentials.

In Ref.~\cite{RM_PRA24}, the ULRM potentials were constructed to be independent of hyperfine coupling. The detailed analysis of the scattering processes now  introduces different ULRM potentials for different hyperfine states of the Fermi superfluid. We have verified that if the two ULRM potentials are artificially set equal, our results reduce to those presented in Ref.~\cite{RM_PRA24} assuming spin-independent ULRM potentials. 
The individual ULRM potentials are truncated to the two outermost wells, shown in Fig.~\ref{Fig:K40Pot}. 
As described in Ref.~\cite{RM_PRA24}, the double-well approximation is indeed physical, because the outermost wells of the ULRM potentials have the highest  probability and Franck-Condon factors for molecule formation. For all the cases presented in the following, we have confirmed that the  single outermost-well and the four-well approximations produce qualitatively  similar features. Henceforth, we only present the results from the double-well approximation.
Moreover, $p$-wave shape resonances \cite{Shaffer2018,PhysRevA.98.042706} in alkali atoms modify the potentials in regions closer to the core and leave unaffected the outermost potentials.

\section{BdG formalism of Rydberg atom-Fermi superfluid systems}\label{Sec:BdG}
The Rydberg atom - Fermi superfluid system is modeled by the following Hamiltonian  
\begin{subequations}
\begin{eqnarray}
\label{Eq:Hint}
\mathcal{H}&=&\mathcal{H}_{BCS}+\sum_{\sigma}\int dx V_{Ryd,\sigma}(x)\psi_\sigma^{\dagger}(x)\psi_\sigma(x) d^\dagger d, ~~~~~~ \\ 
\mathcal{H}_{BCS}&=&\int dx\Big[\sum_{\sigma} \psi_{\sigma}^{\dagger}(x)h_{\sigma}(x)\psi_{\sigma}(x)+ \nonumber \\
& &(\Delta(x) \psi_{\uparrow}^{\dagger}(x)\psi_{\downarrow}^{\dagger}(x)+h.c.)\Big],
\end{eqnarray}
\end{subequations}
following the mean-field BCS-Leggett theory 
\cite{Leggett,Pethick-BEC} for the Fermi superfluid. 
The fermion operator acting on the $\sigma=\uparrow,\downarrow$ component of mass $m$ is $\psi_\sigma$, and $h_\sigma(x)=-\frac{\hbar^2}{2m}\nabla^2+V_{ext,\sigma}(x)-\mu_\sigma$ denotes the single-particle Hamiltonian with $V_{ext,\sigma}(x)$ summarizing the total external confinement. The ULRM potentials $V_{Ryd,\sigma}$ will be implicitly combined with $V_{ext,\sigma}(x)$ in $h_\sigma$. 
As evident from $\mathcal{H}_{BCS}$, direct spin exchange is prohibited.

Here we assume that only a few Rydberg atoms are immersed in a Fermi superfluid background featuring equal populations in the two spin components. Therefore, the background chemical potential of the Fermi superfluid is not affected by the presence of the Rydberg atoms, so that $\mu_\sigma=\mu$ for the bulk chemical potential. However, the introduction of spin-dependent ULRM potentials may induce local population imbalance despite the equal bulk chemical potentials. The gap function is also the order parameter of the Fermi superfluid. Explicitly,   
\begin{align}\label{Eq:DeltaOriginal}
\Delta(x)&=-U\langle\psi_{\downarrow}(x)\psi_{\uparrow}(x)\rangle.
\end{align}
The effective coupling $U<0$ is related 
to the effective 1D scattering length $a_{1D}$~\cite{Olshani,MISTAKIDIS20231} via $U=-\frac{2\hbar^2}{ma_{1D}}$ and it is  controlled by Feshbach resonances~\cite{Cheng_Feshbach}. Since we focus on ground-state properties, 
$\langle \dots \rangle$ denotes the ground-state expectation value at $T=0$. 
As explained in Ref.~\cite{ZwergerBook}, the BCS-BEC crossover occurs when the chemical potential crosses zero, leading to different behavior of the single-particle excitations.

To find the eigenvalues and eigenfunctions of the Hamiltonian $\mathcal{H}$, we implement the BdG formalism \cite{bogoliubov1947theory,BdG-book} to diagonalize it by introducing the canonical transformation 
\begin{eqnarray}
     \psi_{\uparrow,\downarrow}(x)= \sum_{\tilde{n}}[u_{\uparrow,\downarrow}^{\tilde{n}}(x)\gamma_{\tilde{n}}\mp v_{\uparrow,\downarrow}^{\tilde{n}*}(x)\gamma_{\tilde{n}}^{\dagger}]. 
\end{eqnarray}
Here, $u_\sigma^{\tilde{n}}$ and $v_\sigma^{\tilde{n}}$ are the quasiparticle wave functions, which will be determined later. The canonical transformation requires the normalization $\int dx(|u_\sigma^{\tilde{n}}|^2+|v_\sigma^{\tilde{n}}|^2)=1$.
The BdG equation of the considered composite system becomes the following matrix equation:
\begin{equation}\label{eq:subset1}
  \begin{pmatrix}
    h_{\uparrow}& 0 & 0 & \Delta\\
    0 & h^*_{\downarrow} & \Delta  & 0 \\
    0 & \Delta^* & -h_{\uparrow}^* &0 \\
    \Delta^* & 0 & 0 & -h_{\downarrow}^*
    \end{pmatrix}
    \begin{pmatrix}
    u^{\tilde{n}}_{\uparrow}\\
    u^{\tilde{n}}_{\downarrow}\\
    v^{\tilde{n}}_{\uparrow}\\
    v^{\tilde{n}}_{\downarrow}
    \end{pmatrix}
    =E_{\tilde{n}}\begin{pmatrix}
    u^{\tilde{n}}_{\uparrow}\\
    u^{\tilde{n}}_{\downarrow}\\
    v^{\tilde{n}}_{\uparrow}\\
    v^{\tilde{n}}_{\downarrow}
    \end{pmatrix}.
\end{equation}
The dependence on $x$ has been suppressed for brevity.
In the absence of explicit spin-flipping terms that exchange the two spin states, the BdG equation respects a discrete symmetry connecting the positive and negative energy states even when $h_\uparrow$ is different from $h_\downarrow$. Explicitly, the eigen-energy of a state changes sign if $u_\downarrow \leftrightarrow v_\downarrow^*$ and $u_\uparrow \leftrightarrow -v_\uparrow^*$ \cite{BdG-book}. This symmetry property  allows us to examine only the states with positive energies and still acquire the full information.

For the ground state, the gap function shown in Eq.~(\ref{Eq:DeltaOriginal})  becomes
\begin{eqnarray}\label{Eq:Delta}
\Delta(x)=-\frac{U}{2}{\sum_{\tilde{n}}}' [u^{\tilde{n}}_{\uparrow }(x)v^{\tilde{n}*}_{\downarrow}(x)+u^{\tilde{n}}_{\downarrow }(x)v^{\tilde{n}*}_{\uparrow}(x)]. 
\end{eqnarray}
The total fermion density is $n(x)=\sum_\sigma n_{\sigma}(x)$, where      $n_\sigma(x)=\langle\psi_\sigma^\dagger(x)\psi_\sigma(x)\rangle$ results in 
\begin{eqnarray}\label{Eq:neq}
n_{\sigma}(x)={\sum_{\tilde n }}'|v_{\tilde n \sigma}(x)|^2.
\end{eqnarray}
Here, ${\sum_{\tilde{n}}}'$ denotes summation over the positive-energy states. Therefore, only the wave functions $v_{\tilde n \sigma}(x)$ contribute to the ground-state densities. The total particle number of each spin is $N_\sigma=\int n_\sigma(x) dx$, and $N=N_\uparrow + N_\downarrow$. The energy gap for a homogeneous BCS Fermi superfluid is~\cite{Pethick-BEC,Ueda-book} 
\begin{equation}
E_g =
\begin{cases}
\Delta,~\mu > 0,\\
\sqrt{\Delta^2+\mu^2},~\mu < 0. \label{Eq:Eg} \\
\end{cases}       
\end{equation}
In our numerical calculations, we will use the bulk value of $\Delta$ away from the ULRM potentials and the bulk $\mu$ to estimate $E_g$.

In the following, we consider a 1D box of length $L$ with the Rydberg atom placed at the origin ($x=0$) and the dip of its furthest potential well located in the middle ($x/L=1/2$). 
If $R_0$ denotes the distance from the furthest dip of the potential to the Rydberg atom, then $L=2R_0$. The energy unit is chosen as $E_0=\hbar^2/(2mL^2)$. For the 40S (50S)  ULRM potentials of $^{40}$K, $R_0=143$ nm ($197$ nm), so $E_0=1.5$ kHz ($0.6$ kHz).
We discretize the space and implement an iterative method~\cite{BdG-book,PhysRevA.107.063314} to solve the BdG equation until the gap function converges.
In our calculations, we use a grid of $1000$ points which ensures that the observables of interest (described below) remain un-changed 
as the grid is further refined. 
We also introduce the Fermi energy $E_f=\hbar^2 k_f^2/(2m)$ and wavevector $k_f=\pi n_f/2$ where $n_f$ is the density of a noninteracting 1D two-component Fermi gas with equal populations and the same total particle number $N$ in the box as the Fermi superfluid studied here. 


\section{K(50S) ULRM potentials}\label{Sec:Kn50}
\subsection{Population-imbalance transition}
\begin{figure}[t]
\centering
\includegraphics[width=\columnwidth]{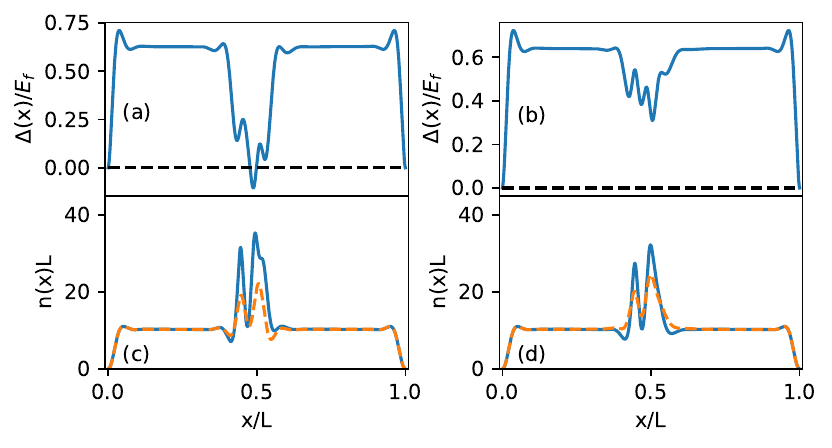}
\caption{Profiles of [(a), (b)] the gap function and [(c), (d)] the densities of the $^{40}$K Fermi superfluid in the presence of K (50S) spin-dependent potentials across the population-imbalance transition on the BCS side. The left (right) column presents the population-imbalanced (equal-population) setting with $U/(E_0 L)=-60$, $\mu/E_0=600$ ($U/(E_0 L)=-61$, $\mu/E_0=600$), resulting in $N_\uparrow=11.5, N_\downarrow=10.5$ ($N_\uparrow=11=N_\downarrow$). The solid and dashed lines in panels (c) and (d) denote the densities  $n_\uparrow$ and  $n_\downarrow$, respectively.}
\label{Fig:n50ProfilesTransition}
\end{figure}

Here we describe the behavior of the composite system according to 
the BdG equation with the double-well approximation of the $^{40}$K (50S) ULRM potentials shown in Fig.~\ref{Fig:K40Pot}(a). 
For these spin-dependent
ULRM potentials, we find a critical point separating the equal-population and population-imbalance cases due to the local polarization potentials as the pairing strength increases. 
In particular, the critical point occurs on the BCS side when the Fermi superfluid has $\mu >0$. Fig.~\ref{Fig:n50ProfilesTransition} presents the gap function and the densities of the components on the two sides of the population-imbalance transition. As one can see, the transition occurs abruptly as the pairing interaction is varied by about 1\%, and the system transforms from equal-population (with $N_\uparrow=N_\downarrow$) to population-imbalance (with $N_\uparrow\neq N_\downarrow$). 
We remark that the population-imbalance transition depends, in general, on the effective coupling $U$, the spin-dependent Rydberg interaction strength $V_{Ryd,\sigma}$, and the chemical potential. For example, in the case of $U=-61E_0$ and $\mu=601E_0$ population imbalance occurs, in contrast to the scenario where $U=-61E_0$ and $\mu=600E_0$ resulting in equal populations.

A careful examination of the density profiles of the components reveals that the equal-population case on the BCS side actually exhibits local population imbalance ($n_\uparrow\neq n_\downarrow$) around the spin-dependent ULRM potentials, see  Fig.~\ref{Fig:n50ProfilesTransition}(d). However, the populations in the entire box are still equal ($N_\uparrow=N_\downarrow$). In contrast, the population-imbalance scenario, illustrated in Fig.~\ref{Fig:n50ProfilesTransition}(c), exhibits both different local densities ($n_\uparrow\neq n_\downarrow$) and different global particle numbers ($N_\uparrow\neq N_\downarrow$).
Meanwhile, the gap function becomes locally negative 
within the ULRM potentials in the presence of global population imbalance with $\Delta N= N_\uparrow-N_\downarrow > 0$ as shown in Fig.~\ref{Fig:n50ProfilesTransition}(a). 
This is reminiscent of the FFLO states with a spatially modulating gap function~\cite{PhysRev.135.A550,LO64}. We will subsequently examine the energy spectrum and eigen-functions from the BdG equation to understand those features.

\subsection{Energy spectrum and wave functions}
\begin{figure}[t]
\centering
\includegraphics[width=0.93\columnwidth]{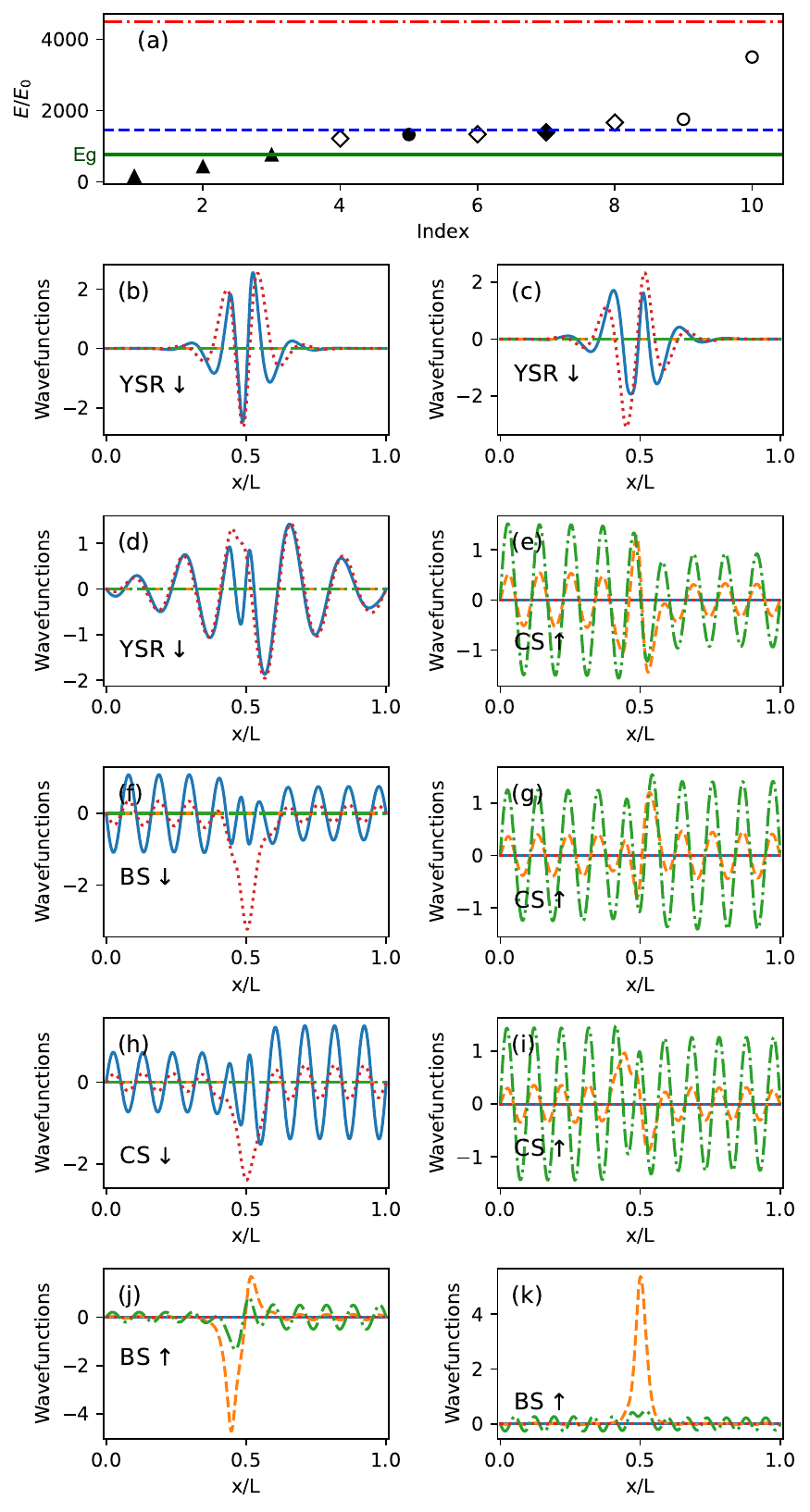}
\caption{Equal-population setup shown in the right column of Fig.~\ref{Fig:n50ProfilesTransition}. 
(a) The energy spectrum of special states. The $n_\downarrow$ ($n_\uparrow$) states are marked by the solid (hollow) symbols while the triangle, circle, and diamond denote the YSR, bound state (BS), and clumpy state (CS), respectively. Also, the energy gap $E_g$ (solid line) and the maximal magnitudes of $V_{Ryd,\downarrow}$ (dashed line) and $V_{Ryd,\uparrow}$ (dot-dash line) are provided. (b)-(k) The wave functions of the states of panel (a) categorized from the lowest to highest energy, where in particular $u_{\uparrow}$, $v_{\uparrow}$, $u_{\downarrow}$, and $v_{\downarrow}$ are represented by the solid, dash, dot-dash, and dotted lines, respectively. 
In all cases, the parameters $U/(E_0 L)=-61$, $\mu/E_0=600$ are considered, resulting in $N_\uparrow=N_\downarrow=11$.}
\label{Fig:BS_Kn50_U61}
\end{figure}

In the following, we focus on special states that are different from typical bulk states and appear on each of the two sides of the population-imbalance transition.
When $V_{Ryd,\uparrow}\neq V_{Ryd,\downarrow}$, the solution of the BdG equation gives four wave functions, namely $u_\uparrow, v_\uparrow, u_\downarrow, v_\downarrow$. While $|v_\sigma|^2$ reflects $n_\sigma$ for $\sigma=\uparrow,\downarrow$ according to Eq.~\eqref{Eq:neq}, the pairing gap has contributions from both $u_\uparrow v_\downarrow^*$ and $u_\downarrow v_\uparrow^*$ according to Eq.~\eqref{Eq:Delta}.
Fig.~\ref{Fig:BS_Kn50_U61}(a) depicts the energy spectrum of special states different from the typical bulk states of the equal-population case near the population-imbalance transition on the BCS side shown in the right panels of Fig.~\ref{Fig:n50ProfilesTransition}. The wave functions of those special states are shown in Fig.~\ref{Fig:BS_Kn50_U61}(b)-(k). The energy gap $E_g$ set by the bulk value of $\Delta$ on the BCS side according to Eq.~\eqref{Eq:Eg} 
corresponds to the solid line in Fig.~\ref{Fig:BS_Kn50_U61}(a), while the maximal depths of $V_{Ryd,\uparrow}$ and $V_{Ryd,\downarrow}$ refer to the dot-dashed and dashed lines, respectively. For the 50S potassium ULRM potentials, the three aforementioned energy scales are $740E_0$, $4500E_0$, and $1500E_0$, respectively, with  $E_0=0.6\textrm{kHz}$.

We have numerically confirmed that the three eigenstates presented in Fig.~\ref{Fig:BS_Kn50_U61}(b)-(d) have $v_\uparrow=0$ and therefore only contribute to $n_\downarrow$ according to Eq.~\eqref{Eq:neq}. Since those states have overlapping $v_\downarrow$ and $u_\uparrow$, they contribute to the gap function according to Eq.~\eqref{Eq:Delta}. Importantly, their energies lie below the energy gap $E_g$, as it can be seen from the three lowest triangles that are located below the solid line in Fig.~\ref{Fig:BS_Kn50_U61}(a). However, we remark that these $n_\downarrow$ states are not simply confined by  $V_{Ryd,\downarrow}$ but importantly by the polarization potential $V_{Ryd,\uparrow}-V_{Ryd,\downarrow}$ created by the difference between the two spin-dependent ULRM potentials. We have verified that similar in-gap states still emerge in an artificial case when $V_{Ryd,\uparrow}$ is the same but $V_{Ryd,\downarrow}=0$. 
Therefore, those in-gap states are actually the YSR states since the polarization potential pulls down the energies of the Cooper pairs, thereby leaving the YSR states below $E_g$, as shown by the triangles below the solid line in Fig.~\ref{Fig:BS_Kn50_U61}(a). 
To further testify the origin of the in-gap YSR states, we have artificially set $V_{Ryd,\downarrow}=V_{Ryd,\uparrow}$ in the BdG calculation and confirm that 
they disappear since the 
polarization potential no longer exists.

It is worth noticing  that the scattering of the Rydberg electron with the ground state fermions lying in the hyperfine states of the Fermi superfluid produces the electron spin-dependent ULRM potentials. However, the Rydberg impurity itself does not contribute any  additional spin degree of freedom into the molecular formation. For the Rydberg atom-Fermi superfluid system, the spin-dependent ULRM potentials induce a local magnetic polarization (effective Zeeman splitting) to form the YSR states. This is in contrast to superconductor-impurity systems where a magnetic impurity carries spin and may form spin-singlet or spin-triplet bound states with the Fermi superfluid due to the combination of the impurity and the bound fermions.

\begin{figure}[t]
\centering
\includegraphics[width=0.95\columnwidth]{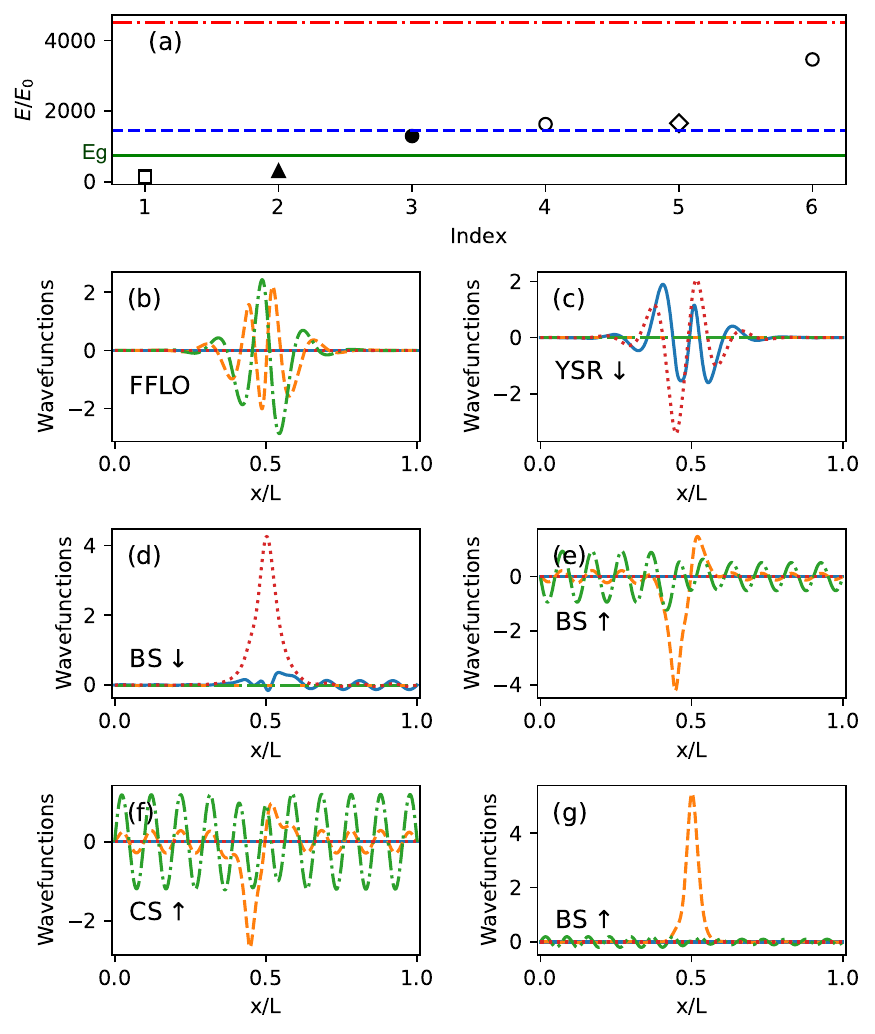}
\caption{Population-imbalanced setting of the left column of Fig.~\ref{Fig:n50ProfilesTransition}. 
(a) Energy spectrum of special states. The $n_\downarrow$ ($n_\uparrow$) states are designated by the solid (hollow) symbols while the square, triangle, circle, and diamond refer to the FFLO-like, YSR, BS, and CS states, respectively. The energy gap $E_g$ (solid line) and the maximal magnitudes of $V_{Ryd,\downarrow}$ (dash line) and $V_{Ryd,\uparrow}$ (dot-dash line) are also shown. (b)-(g) The wave functions of the states presented in panel (a) from the lowest to the highest energy, showing $u_{\uparrow}$ (solid lines), $v_{\uparrow}$ (dash lines), $u_{\downarrow}$ (dot-dash lines), and  $v_{\downarrow}$ (dotted lines).   
The parameters are $U/(E_0 L)=-60$, $\mu/E_0=600$, resulting in $N_\uparrow=11.5$ and  $N_\downarrow=10.5$.
}
\label{Fig:BS_Kn50_U60}
\end{figure}

Above the energy gap, $E_g$, of the Fermi superfluid, the states with higher densities around the ULRM potentials can be categorized into two groups, the bound states and clumpy states, according to their energies and wave function structure. Also, we can identify the density contributions from the wave functions according to Eq.~\eqref{Eq:neq}. Inspecting  Fig.~\ref{Fig:BS_Kn50_U61}(f) and (h) it becomes evident that they refer to $n_\downarrow$ states since $v_\uparrow=0$, while panels (e), (g), (i)-(k) correspond to $n_\uparrow$ states due to $v_\downarrow=0$. We notice that in general, the  spin-$\uparrow$ localized states possess larger binding energies compared to the spin-$\downarrow$ ones 
since $V_{Ryd,\uparrow}$ is deeper than $V_{Ryd,\downarrow}$, see also the hollow symbols in Fig.~\ref{Fig:BS_Kn50_U61}(a). By examining the spatial localization behavior of the ensuing states, it is possible to discern that the one depicted in  Fig.~\ref{Fig:BS_Kn50_U61}(f) is a bound state trapped by the $V_{Ryd,\downarrow}$ potential, while the ones illustrated in panels (j) and (k) are bound states within the $V_{Ryd,\uparrow}$ potential. 
This is due to the fact that they exhibit prominent peaks within the corresponding potential and their energies do not exceed the depth of the respective spin-dependent ULRM potentials, as shown in Fig.~\ref{Fig:BS_Kn50_U61}(a). Nevertheless, there is an additional group of states, which we dub clumpy states, depicted in Fig.~\ref{Fig:BS_Kn50_U61}(e), (g), (h), (i). The wave function structure of these clumpy states is arguably more homogeneous and oscillatory 
in real space as compared to the other states, and hence they are not as localized as the YSR states or the bound states. Moreover, the clumpy states are energetically higher-lying 
on the spectrum and 
have higher density contributions around the ULRM potentials. 
Therefore, they balance the spin populations 
in order to reduce the 
imbalance in the presence of the local spin-polarization potential. We note that the clumpy state of panel (h) may be hybridized with the bound state of panel (f) to have a similar but more spread-out shape. 
To differentiate the above-discussed three types of states, we use the triangle, circle, and diamond symbols marking the YSR, bound, and clumpy states in Fig.~\ref{Fig:BS_Kn50_U61}(a), respectively.

A slight decrease of the pairing strength leads to the population-imbalanced case on the BCS side shown in the left column of Fig.~\ref{Fig:n50ProfilesTransition}.
The underlying energy spectrum is provided in Fig.~\ref{Fig:BS_Kn50_U60}(a) along with the eigenstate wave functions being different from the typical bulk ones shown in Fig.~\ref{Fig:BS_Kn50_U60}(b)-(g).  
It is apparent that the major difference from the equal-population case is the emergence of an additional class of spatially localized states. 
Indeed, for the energetically lowest localized state presented in Fig.~\ref{Fig:BS_Kn50_U60}(b), the wave functions $v_\uparrow$ and $u_\downarrow$ are out-of-phase, rendering a negative contribution to the gap function according to Eq.~\eqref{Eq:Delta}. 
This explains the dipping of the gap function below zero inside the ULRM potentials which can be seen in Fig.~\ref{Fig:n50ProfilesTransition}(a). 

Spatial modulations of the gap function in population imbalanced Fermi superfluids with a uniform polarization potential 
have been discussed in the context of FFLO states~\cite{PhysRev.135.A550,LO64}. In Appendix~\ref{Sec:AppA}, we demonstrate some properties of a FFLO superfluid in a box potential induced by a uniform polarization field. Here, we find  that the spin-dependent ULRM potentials of the Rydberg atom-Fermi superfluid system give rise to  
a local polarization potential. The latter can induce a sign change in the gap function within the potential on the population-imbalance side. Moreover, the out-of-phase wave functions presented in Fig.~\ref{Fig:BS_Kn50_U60}(b) provide a concise explanation of the negativity of the gap function in the region where the polarization potential is maximal. 
Given the presence of local population imbalance and the accompanied sign change of the gap function (reminiscent of the FFLO states induced by a homogeneous polarization field shown in Appendix~\ref{Sec:AppA}), we call this class of states FFLO-like and emphasize that their effects are confined by the local polarization potential. We also remark that the lowest-energy FFLO-like state here contributes to $n_\uparrow$ because $v_\downarrow=0$, which is opposite to the lowest-energy YSR state contributing to $n_\downarrow$ in the adjacent equal-population case. 
This property provides a further distinction between the FFLO-like states and the previously discussed YSR ones. 

For clarity, the square symbol is used to denote the FFLO-like state in Fig.~\ref{Fig:BS_Kn50_U60}(a). 
We emphasize that a typical FFLO state occurs in a homogeneous polarization potential and represents a global phase of matter of a population imbalanced superfluid. Typical signatures of the FFLO state include i) a modulated order parameter that change signs in real space, ii) out-of-phase wave functions where the gap function changes signs, and iii) population imbalance accompanying the sign-change of the order parameter, as illustrated in Appendix~\ref{Sec:AppA}. In contrast,  the spin-dependent ULRM potentials correspond to an effective local magnetic polarization, i.e., Zeeman splitting, so the FFLO-like state is a local effect when the Fermi superfluid becomes spin-polarized within the local potential. Importantly, the sign-change of the order parameter in real space is shown in Fig.~\ref{Fig:n50ProfilesTransition}(a), accompanied by population imbalance in Fig.~\ref{Fig:n50ProfilesTransition}(b) and out-of-phase wave functions in Fig.~\ref{Fig:BS_Kn50_U60}(b). These properties corroborate the notion of the FFLO-like states in the Rydberg atom - Fermi superfluid systems. 

The second lowest-energy state illustrated in Fig.~\ref{Fig:BS_Kn50_U60}(c) corresponds to a $n_\downarrow$ state because its $v_\uparrow =0$.
Although the participating $v_\downarrow$ and $u_\uparrow$ are slightly out-of-phase, this state does not contribute negatively to the gap function according to Eq.~\eqref{Eq:Delta}. Therefore, it is a YSR state with its energy lying below the energy gap of the Fermi superfluid, see Fig.~\ref{Fig:BS_Kn50_U60}(a). 
Above the energy gap, $E_g$, designated by the solid line in Fig.~\ref{Fig:BS_Kn50_U60}(a), we demonstrate the bound states in $V_{Ryd,\downarrow}$ and $V_{Ryd,\uparrow}$ in Fig.~\ref{Fig:BS_Kn50_U60}(d) and (g), respectively. 
The eigenstates presented in Fig.~\ref{Fig:BS_Kn50_U60}(e) and (f) suggest that the second bound state in $V_{Ryd,\uparrow}$ may be hybridized with a clumpy state forming two states with similar wave functions but slightly different energies. 
We remark that the total atom number difference $\Delta N = N_\uparrow-N_\downarrow=1$ in this case, and the major contribution is from the emergence of the FFLO-like state since it contributes to $n_\uparrow$ and is absent in the equal-population case.


All the special states presented in Figs.~\ref{Fig:BS_Kn50_U61} and \ref{Fig:BS_Kn50_U60} being different from typical bulk states refer to single-particle eigenstates in the sense that only a fermion of a specific spin state from a broken Cooper pair is trapped by the ULRM potentials. 
Therefore, the bound states represent diatomic Rydberg molecules since only a fermion is trapped by the Rydberg atom. When summing up the populations of the fermions, we find that the FFLO-like state accompanies the breaking of population balance and leads to a locally polarized Fermi superfluid with a sign-change of the gap function inside the polarization potential. 
It is also worth noting that it is possible to produce $N_\uparrow-N_\downarrow > 1$ in the BdG calculations with lower fermion densities by further adjusting the chemical potential and pairing strength of the Fermi superfluid, but those higher-polarization states 
exhibit more dramatic spatial modulations and noisy behavior.

\section{K(40S) ULRM potentials}\label{Sec:Kn40}
\subsection{Population-imbalance transition}
Next, we consider stronger spin-dependent ULRM potentials to probe the physics on the BEC side of the Fermi superfluid. The spin-dependent ULRM potentials of the (40S) state of a potassium atom with the two hyperfine states of a $^{40}$K atomic Fermi superfluid are shown in Fig.~\ref{Fig:K40Pot}(b) along with their double-well approximations. The depths of the potential wells are deeper and the furthest well is closer to the core compared to those from the (50S) Rydberg state.
Due to the deeper ULRM potentials of the (40S) Rydberg state, the transition from the equal-population to population-imbalanced cases of the Fermi superfluid is pushed to the BEC side with $\mu>0$, in contrast to the population-imbalance transition  occurring on the BCS side in the presence of the shallower (50S) ULRM potentials.

\begin{figure}[t]
\centering
\includegraphics[width=\columnwidth]{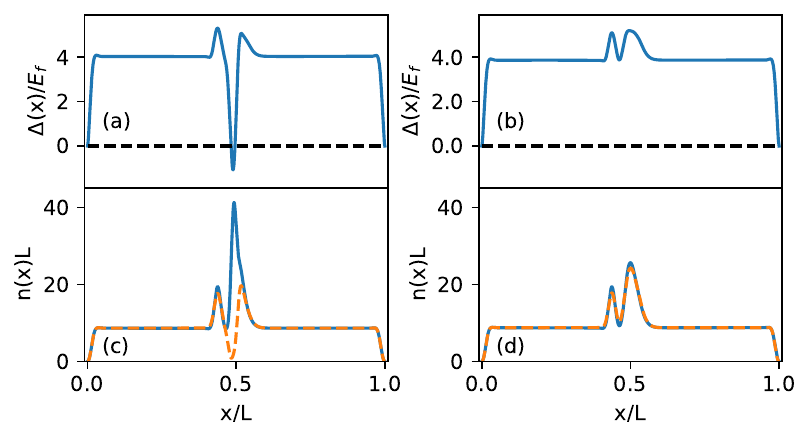}
\caption{Profiles of [(a), (b)] the gap function  and [(c), (d)] the densities of a $^{40}$K Fermi superfluid in the presence of K(40S) spin-dependent ULRM potentials across the population-imbalance transition on the BEC side. The left (right) column depicts the population-imbalanced (equal-population) setup characterized by  $U/(E_0 L)=-150$, $\mu/E_0=-800$ (($U/(E_0 L)=-151,\mu/E_0=-800$), and leading to populations  $N_\uparrow=9.8, N_\downarrow=8.8$ ($N_\uparrow=N_\downarrow=9.6$). The solid and dashed lines in panel (c) and (d) refer to the densities  $n_\uparrow$ and $n_\downarrow$ of the Fermi superfluid, respectively.}
\label{Fig:ProfilesTransition}
\end{figure}

Fig.~\ref{Fig:ProfilesTransition} contrasts two cases with less than $1\%$ variation of the pairing strength under the double-well approximations of the (40S) ULRM potentials. The Fermi superfluid exhibits equal populations on one side and population imbalance on the other. The transition takes place abruptly as the pairing interaction changes, just like the (50S) case in Sec.~\ref{Sec:Kn50}. 
In contrast to the equal-population scenario depicted in Fig.~\ref{Fig:n50ProfilesTransition} on the BCS side under the (50S) ULRM potentials exhibiting local population imbalance ($n_\uparrow\neq n_\downarrow$)
but overall equal populations ($N_\uparrow=N_\downarrow$), the equal-population case under the (40S) ULRM potentials shown in Fig.~\ref{Fig:ProfilesTransition}(d) on the BEC side 
exhibits equal populations throughout the system with $n_\uparrow=n_\downarrow$.
Across the population-imbalance transition, the density profiles of Fig.~\ref{Fig:ProfilesTransition} (c) feature population imbalance with both $n_\uparrow\neq n_\downarrow$ and $N_\uparrow\neq N_\downarrow$ while the gap function changes sign within the spin-dependent ULRM potentials as can be seen in Fig.~\ref{Fig:ProfilesTransition}(a). 
The presence of population imbalance and sign-change of the gap function indicate once more (as we discuss below) that local FFLO-like states are created by the strong polarization potential but this time 
on the BEC side.

The population-imbalance transition can also be observed by slightly varying the chemical potential of the Fermi superfluid with the (40S) ULRM potentials as well. For instance, the case with $U=-150E_0$ and $\mu=-801E_0$ exhibits equal populations while the case of $U=-150E_0$ and $\mu=-800E_0$ features population imbalance. The sensitivity of the population-imbalance transition to the parameters $U$ and $\mu$ occurs since the transition takes place  in a  localized spatial   region which is influenced by the spin-dependent ULRM potentials. 
Therefore, tuning the  global system parameters may effectively alter the local states.

\subsection{Energy spectrum and wave functions}
\begin{figure}[t]
\centering
\includegraphics[width=0.95\columnwidth]{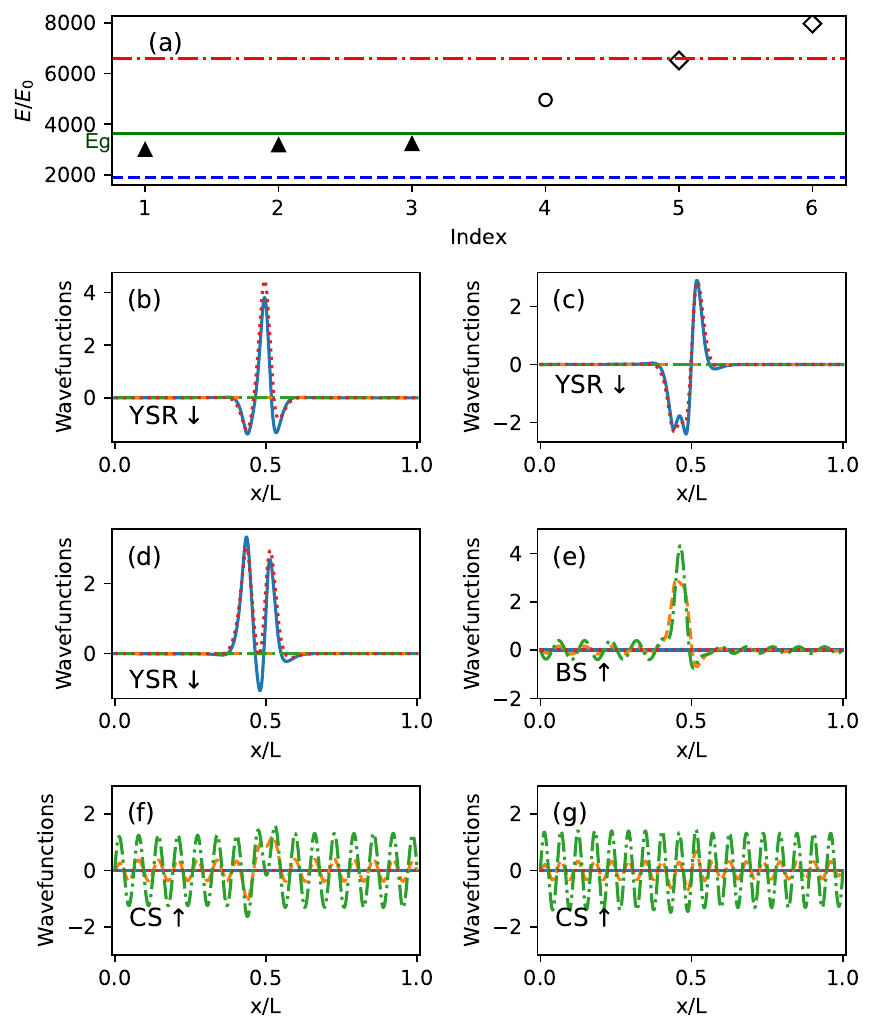}
\caption{Equal-population setup on the BEC side corresponding to the right column of Fig.~\ref{Fig:ProfilesTransition}. 
(a) Energy spectrum of special states. The solid (hollow) symbols refer to states contributing to $n_\downarrow$ ($n_\uparrow$) while the triangle, circle, and diamond signal the YSR, bound state (BS), and clumpy state (CS), respectively. The solid line marks the energy gap $E_g$, while the dash and dot-dash lines are the maximal magnitudes of $V_{Ryd,\downarrow}$ and $V_{Ryd,\uparrow}$, respectively. (b)-(g) The wave functions of the states in panel (a) from the lowest to the highest energy. The eigenfunctions $u_{\uparrow}$, $v_{\uparrow}$, $u_{\downarrow}$, $v_{\downarrow}$ are represented by the solid, dash, dot-dash, and dotted lines, respectively. 
In all cases, $U/(E_0 L)=-151$, $\mu/E_0=-800$, resulting in $N_\uparrow=N_\downarrow=9.6$.}
\label{Fig:BS_K40_U151}
\end{figure}

The energy spectrum of eigenstates regarding the equal-population case [see also the right column of Fig.~\ref{Fig:ProfilesTransition}] on the BEC side near the transition, which are different from the typical bulk states, is provided in Fig.~\ref{Fig:BS_K40_U151}(a) together with their wave functions in the rest of the panels. 
The energy gap $E_g$ set by the bulk value of $\sqrt{\mu^2+\Delta^2}$ on the BEC side according to Eq.~\eqref{Eq:Eg} is denoted by the solid line in Fig.~\ref{Fig:BS_K40_U151}(a) while the dot-dash and dashed lines mark the maximal depths of $V_{Ryd,\uparrow}$ and $V_{Ryd,\downarrow}$ respectively. For the (40S) potassium ULRM potentials, the three aforementioned energy scales are $3600E_0$, $6600E_0$, and $1900E_0$, respectively, where $E_0=1.5\textrm{kHz}$.

From our previous discussion we can readily infer that the three lowest-energy states in Fig.~\ref{Fig:BS_K40_U151}(b)-(d) are YSR states located below the energy gap $E_g$ since their energies are below the solid line in panel (a). Moreover, their vanishing $v_\uparrow$ indicating they only contribute to $n_\downarrow$ since $V_{Ryd,\uparrow} > V_{Ryd,\downarrow}$. For higher-energy states with higher densities around the ULRM potentials, there are again two classes. The state presented in panel (e) corresponds to the lowest bound state within  $V_{Ryd,\uparrow}$. Since $E_g$ is larger than $V_{Ryd,\downarrow}$, see also panel (a), 
we do not find any bound states around $V_{Ryd,\downarrow}$ other than the YSR ones. This is different from the (50S) case on the BCS side shown in Fig.~\ref{Fig:BS_Kn50_U61} where $E_g < V_{Ryd,\downarrow}$, so there is room for bound states in $V_{Ryd,\downarrow}$ with energies above those of the YSR states.
On the other hand, there are again energetically higher-lying clumpy states, see panels (f) and (g), which are crucial for maintaining equal spin populations of the Fermi superfluid throughout its spatial extent. 
For clarity, we note that the triangle, circle, and diamond symbols in  Fig.~\ref{Fig:BS_K40_U151}(a) refer to  the YSR, bound, and clumpy states  respectively.

\begin{figure}[t]
\centering
\includegraphics[width=0.95\columnwidth]{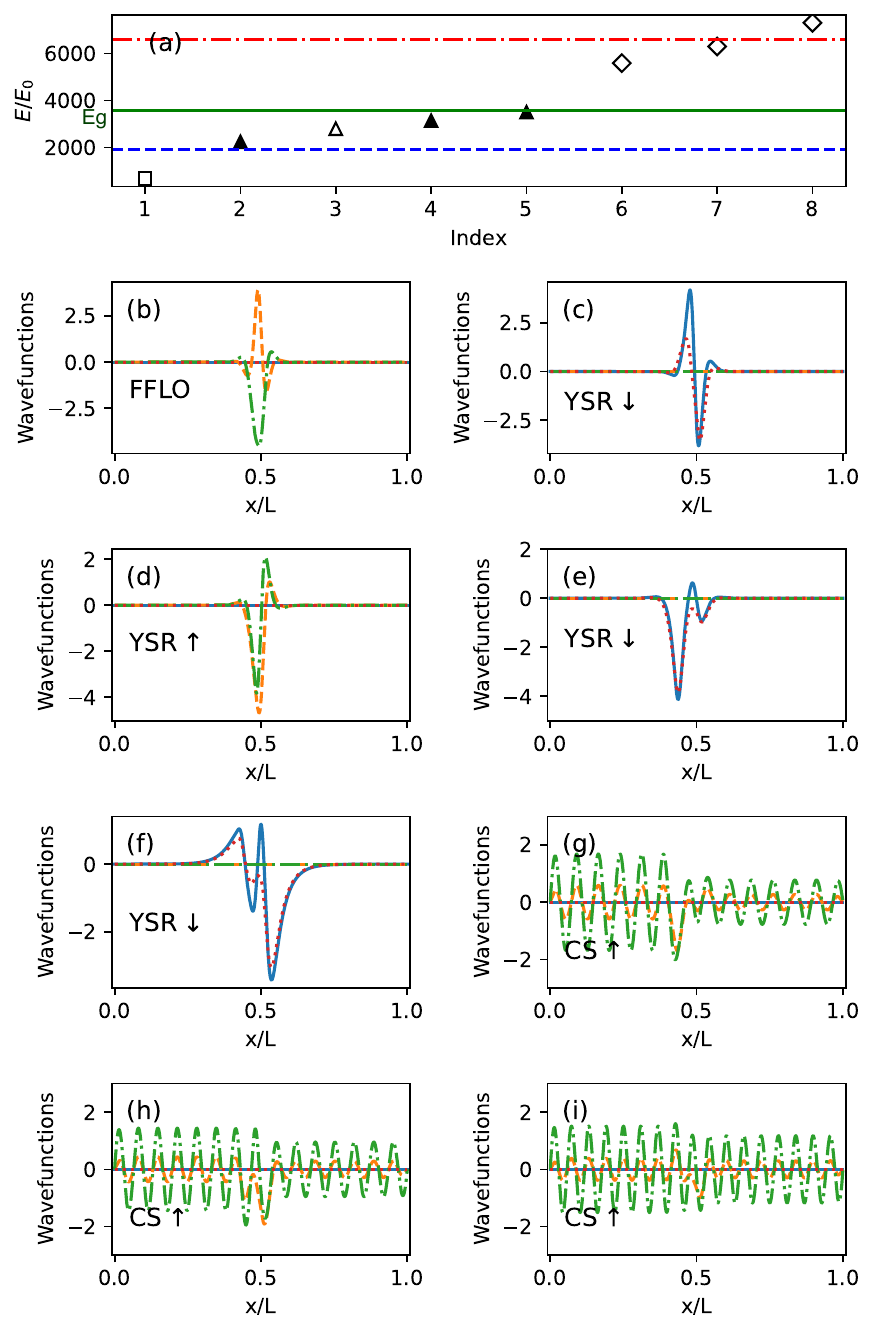}
\caption{Population-imbalanced case on the BEC side shown in the left column of Fig.~\ref{Fig:ProfilesTransition}. 
(a) Energy spectrum of special states. Contributions to $n_\downarrow$ ($n_\uparrow$) are marked by solid (hollow) symbols. Also, the square, triangle, circle, and diamond represent the FFLO-like, YSR, BS, and the CS states. 
The solid line illustrates the energy gap $E_g$, while the dash and dot-dash lines show the maximal magnitudes of $V_{Ryd,\downarrow}$ and $V_{Ryd,\uparrow}$, respectively. 
Panels (b)-(i) depict the wave functions of the states of panel (a) classified from the lowest to the highest energy. The eigenfunctions $u_{\uparrow}$,  $v_{\uparrow}$, $u_{\downarrow}$, and $v_{\downarrow}$ are denoted by the solid, dash, dot-dash, and dotted lines, respectively. 
The parameters are $U/(E_0 L)=-150$ and $\mu/E_0=-800$, yielding $N_\uparrow=9.8$, and $N_\downarrow=8.8$.}
\label{Fig:BS_K40_U150}
\end{figure}

On the other hand, Fig.~\ref{Fig:BS_K40_U150} presents the energy spectrum in panel (a) and the respective wave functions of those special eigenstates different from the typical bulk ones for the population-imbalanced setting of the left column of Fig.~\ref{Fig:ProfilesTransition}. 
Again, the solid, dot-dash, and dash lines in panel (a) indicate $E_g$, maximal $V_{Ryd,\uparrow}$, and maximal $V_{Ryd,\downarrow}$, respectively. Here, the lowest-energy state whose wave functions are shown in panel (b) is a FFLO-like state featuring out-of-phase wave functions between $u_\downarrow$ and $v_\uparrow$. 
This contributes negatively to the gap function according to Eq.~\eqref{Eq:Delta}. The emergence of the FFLO-like state in the presence of population imbalance explains how the gap function turns to be negative within the spin-polarized ULRM potentials. Moreover, the FFLO-like state contributes to $n_\uparrow$ according to Eq.~\eqref{Eq:neq} since $v_\downarrow =0$, which is different to the lowest-energy YSR state contributing to $n_\downarrow$. 
We use the square symbol to denote the FFLO-like state in Fig.~\ref{Fig:BS_K40_U150}(a).

In a homogeneous polarization field, the 1D FFLO state is expected to survive in a broad range when population imbalance is present~\cite{PhysRevResearch.3.043105}. This implies a possible existence of the local FFLO-like state in the BEC regime. However, we remark that the results presented here are not in the deep BEC regime because the bound state shown in Fig.~\ref{Fig:BS_K40_U151} (e) indicates the spin-dependent ULRM potentials can break Cooper pairs and probe fermionic physics. Also, due to the relatively small spatial width of the ULRM potentials, the local FFLO-like state does not exhibit modulations within the potential. Rather, the order parameter changes sign in the FFLO-like state, similar to the sign-change of the order parameter near the hard-wall boundaries in the FFLO state with a homogeneous polarization field shown in Appendix~\ref{Sec:AppA}. Moreover, there are multiple out-of-phase wave functions in the FFLO state with a homogeneous polarization field but only one FFLO-like state with out-of-phase wave function in the Rydberg-Fermi superfluid system.

The next four lowest-energy states shown in Fig.~\ref{Fig:BS_K40_U150}(c)-(f) refer to YSR states because their energies are below the energy gap $E_g$, see also Eq.~\eqref{Eq:Eg}. We notice that the eigenstates of panel (d) contribute to $n_\uparrow$ because its $v_\downarrow=0$, while the eigenstates of panels (c), (e), and (f) with $v_\uparrow=0$ are related to $n_\downarrow$ in line with Eq.~\eqref{Eq:neq}. Interestingly, this is in contrast to the adjacent equal-population setup illustrated in Fig.~\ref{Fig:BS_K40_U151}, where the YSR states are all spin-$\downarrow$ states because their $v_\uparrow=0$. At higher energies, i.e., above the energy gap $E_g$, we only find clumpy states presented in panels (g)-(i) which offset the populations but as before characterized by a relatively less prominent localization behavior. The spin population difference is $N_\uparrow-N_\downarrow=1$, whose major contribution stems from the FFLO-like state.

There is also an interesting distinction between the bound states of the spin-dependent ULRM potentials depicted in Fig.~\ref{Fig:BS_K40_U151} for the equal-population case and those in the presumably spin-independent ULRM potential ($V_{Ryd,\uparrow}=V_{Ryd,\downarrow}$) studied in the work of Ref.~\cite{RM_PRA24} on the BEC side. In particular, when the ULRM potentials of the two spins are equal and deep, it is possible to trap tightly-bound Cooper pairs on the BEC side to form a triatomic Rydberg molecule. The trapped Cooper pairs will appear in the BdG eigenstates as a pair of twin states with almost degenerate eigenvalues and similar wave functions but of opposite spins. In contrast, Fig.~\ref{Fig:BS_K40_U151} does not unveil any such twin states, which in turn implies that only diatomic Rydberg molecules are formed. We have verified this mechanism by artificially setting $V_{Ryd,\uparrow}=V_{Ryd,\downarrow}$ in the BdG equation, and identifying that there are indeed twin bound states of opposite spins from a trapped Cooper pair on the BEC side similar to those reported in Ref.~\cite{RM_PRA24}.

The absence of a tightly bound Cooper pair trapped by the potassium (40S) spin-dependent ULRM potentials on the BEC side is due to a combination of the following physical mechanisms. Firstly, the energy gap $E_g$ on the BEC side is relatively large compared to the smaller ULRM potential $V_\downarrow$ as shown in Fig.~\ref{Fig:BS_K40_U151}(a). 
This renders the two ULRM potentials unlikely to cooperate and trap a Cooper pair together since the pair-breaking energy $E_g$ is in-between them. Secondly, the emergence of the YSR states at low energies favor $n_\downarrow$ within the spin-dependent ULRM potentials. Meanwhile, there are higher-energy clumpy states with higher densities around the ULRM potentials to compensate for the local population imbalance, thereby disfavoring the trapping of Cooper pairs with equal spin contributions at nearby energies. Although our results do not rule out the possibility of forming triatomic Rydberg molecules in strongly spin-dependent ULRM potentials, future studies are required for identifying sufficiently deep 
ULRM potentials for both spins and fine-tune the pairing strength of the Fermi superfluid to fit the energy gap $E_g$ below both ULRM potentials even on the BEC side with a relatively large gap function.

\section{Experimental implications}\label{Sec:Exp}
We found that the population-imbalance induced by the ULRM potentials typically favors smaller pairing strength $U$, higher chemical potential $\mu$, and stronger local polarization potential $V_{Ryd,\uparrow}-V_{Ryd,\downarrow}$.
Local measurements of the pairing gap of atomic Fermi superfluids can be achieved through spatially resolved radio-frequency (rf) spectroscopy~\cite{PhysRevLett.99.090403,doi:10.1126/science.aan5950}. 
The local modulations of the gap function by the spin-dependent ULRM potentials demonstrated in Figs.~\ref{Fig:n50ProfilesTransition} and \ref{Fig:ProfilesTransition}, indicate the creation of intriguing eigenstates.  
Additionally, the sign-change behavior of the gap function across the population-imbalance transition is associated with nodal points where the gap vanishes. Identifying those nodal points then implies the emergence of the local FFLO-like states. 
For instance, adding Rydberg excitations to the setup discussed in Ref.~\cite{Liao2010} would enable the  observation of quasi-1D FFLO-like states in population-imbalanced atomic Fermi superfluids,  
and reveal the density modulations when state-dependent imaging of the different components in the Fermi superfluid is employed. 
However, a direct measurement of the relative phase of the pairing gap for further characterizing the   
FFLO-like states remains a challenge.

To experimentally infer the formation of the in-gap YSR and FFLO-like states induced by the spin-dependent ULRM potentials, it is possible to utilize molecular line spectroscopy~\cite{PhysRevLett.120.083401,PhysRevA.97.022707,Exner25}. 
This will reveal the binding energies associated with the aforementioned localized states. We recall that the lowest-energy state is an FFLO-like state in the presence of population imbalance with the YSR states forming below the energy gap $E_g$, given by Eq.~\eqref{Eq:Eg} and set by the bulk values. 
Additional bound states in each of the ULRM potentials will contribute to the excitation spectrum of the Rydberg-molecule. Since the higher-energy clumpy states are not genuine localized states, their detection is anticipated to be more challenging but their presence for compensating the densities around the ULRM potentials could be observable from the state-resolved imaging of the density profiles of the Fermi superfluid~\cite{PhysRevA.77.033401,PhysRevA.97.023410,Liao2010}.

Furthermore, the YSR states are present regardless of whether the spin-dependent potentials are attractive or repulsive because it is the polarization potential constructed by the difference of the potentials that matters. However, repulsive potentials typically studied in solid-state systems~\cite{RevModPhys.78.373,HEINRICH20181} lead to a suppression of the density profiles around the potential in addition to the decaying gap function around the polarization potential. In contrast, the attractive ULRM potentials introduce states with higher densities around the potentials while the gap function exhibits modulations. Moreover, the YSR states in the attractive spin-dependent ULRM potentials may not be the lowest-energy states since the local FFLO-like state in the population-imbalance regime can have even lower energy.

\section{Summary \& Outlook}\label{Sec:Conclusion}
We explored the energy spectrum and eigenstates of a two-component atomic Fermi superfluid trapped in spin-dependent ULRM potentials created by K(50S) and K(40S) isotopic Rydberg excitations. 
These different isotopic excitations provide access to physics on the BCS and BEC sides of the Fermi
superfluid, respectively. 
Based on the BdG formalism, we predict  
a transition from equal-population to population imbalance since the difference of the ULRM potentials corresponds to a local polarization potential. 
In addition, a plethora of interesting eigenstates was identified and subsequently analyzed, 
including local FFLO-like states with out-of-phase wave functions contributing to population imbalance as well as a sign change of the gap function around the spin-dependent ULRM potentials as the system crosses the aforementioned transition. 
Moreover, low-energy YSR states lying within the energy gap, spin-polarized bound states by the corresponding ULRM potentials, as well as clumpy states at higher energies which compensate for the density difference, are observed as eigenstates within the BdG spectrum. 
The Rydberg atom-Fermi superfluid composite system thus allows us to explore competitions between Rydberg-molecule formation, Cooper pairing, and local spin-polarization in interacting multi-component quantum many-body systems.

An interesting avenue for further studies is to dynamically probe the formation of the YSR and FFLO-like states, by a spin-dependent Rydberg quench and monitor the generation of these localized states. 
Numerical simulations of rf spectroscopy or pump-probe techniques to identify the resonances corresponding to these states or admixtures thereof are certainly of interest. 
Furthermore, understanding the correlated character of various types of localized states using sophisticated numerical methods constitutes another intriguing perspective. 

Moreover, quantum gas microscopy~\cite{Gierling2011,10.1093/nsr/nww023,Stecker_2017}, currently in lattice systems, can be emulated in trapped quantum gases with Rydberg molecules to probe spatial correlations \cite{whalen2019,hollerith2019}  associated with pairing or localization. A combination of those measurement techniques may further elucidate interesting properties of the special states discussed herein for atomic Fermi superfluids trapped in spin-dependent ULRM potentials.


\begin{acknowledgments} 
C. C. C. was partly supported by the NSF (No. PHY-2310656). Support for ITAMP by the NSF is acknowledged. 
S.I.M acknowledges support from the Missouri University of Science and
Technology, Department of Physics, Startup fund. 
\end{acknowledgments}

\appendix

\section{FFLO states in a homogeneous polarization potential}\label{Sec:AppA}

\begin{figure}[t]
\centering
\includegraphics[width=\columnwidth]{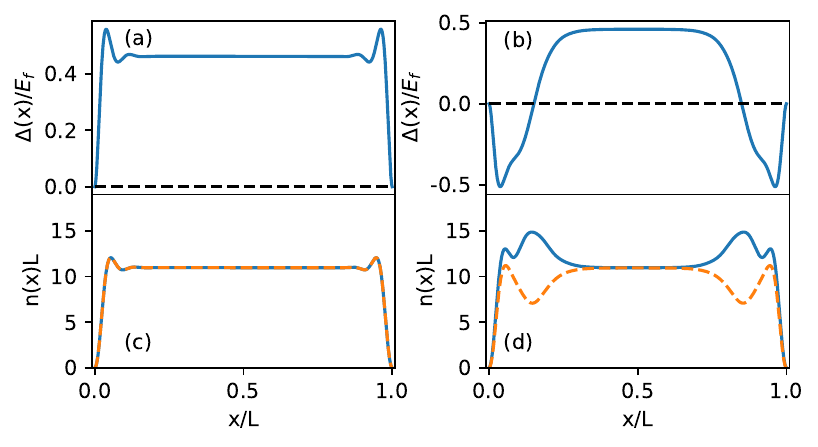}
\caption{Profiles of [(a), (b)] the gap function and [(c), (d)] spin densities extracted from the BdG equation for   $\mu_\downarrow=\mu_\uparrow -h$ and in the absence of ULRM potentials. The uniform polarization potential is denoted by $h$.  The left (right) panels depict the equal-population (population-imbalance) scenario with $U/(E_0 L)=-50, \mu_\uparrow/E_0=1000,h/E_0=670$, leading to $N_\uparrow=10.5=N_\downarrow$ ($U/(E_0 L)=-50, \mu_\uparrow/E_0=1000,h/E_0=675$, resulting in  $N_\uparrow=11.5,N_\downarrow=9.5$). The emergence of the FFLO state near the two hard walls of the box is evident in the right panels.}
\label{Fig:Profiles_Homo}
\end{figure}

To further justify the FFLO-like notion in in Figs.~\ref{Fig:BS_Kn50_U60} and \ref{Fig:BS_K40_U150}, we next study the existence of FFLO states 
in a 1D box within the BdG formalism. 
For this investigation, the ULRM potentials are dropped and a homogeneous spin-polarization potential $h$ is introduced leading to $\mu_\downarrow=\mu_\uparrow -h$ in the BdG equation. The box size $L$ is chosen to be consistent with that in the main text. 
Also in this system a transition from equal-population to population-imbalance takes place as $h$ increases. Specifically, Fig.~\ref{Fig:Profiles_Homo} shows the gap function and the spin density profiles for two characteristic  cases on the opposite sides of the population-imbalance transition induced by $h$. 
Notice also that the particle number difference is a multiple of $2$ as $h$ increases. It can be readily seen that the gap function changes signs around the two hard-wall boundaries. The modulation of the gap function is an indication of the emergence of a FFLO superfluid with a modulating order parameter.

\begin{figure}[t]
\centering
\includegraphics[width=\columnwidth]{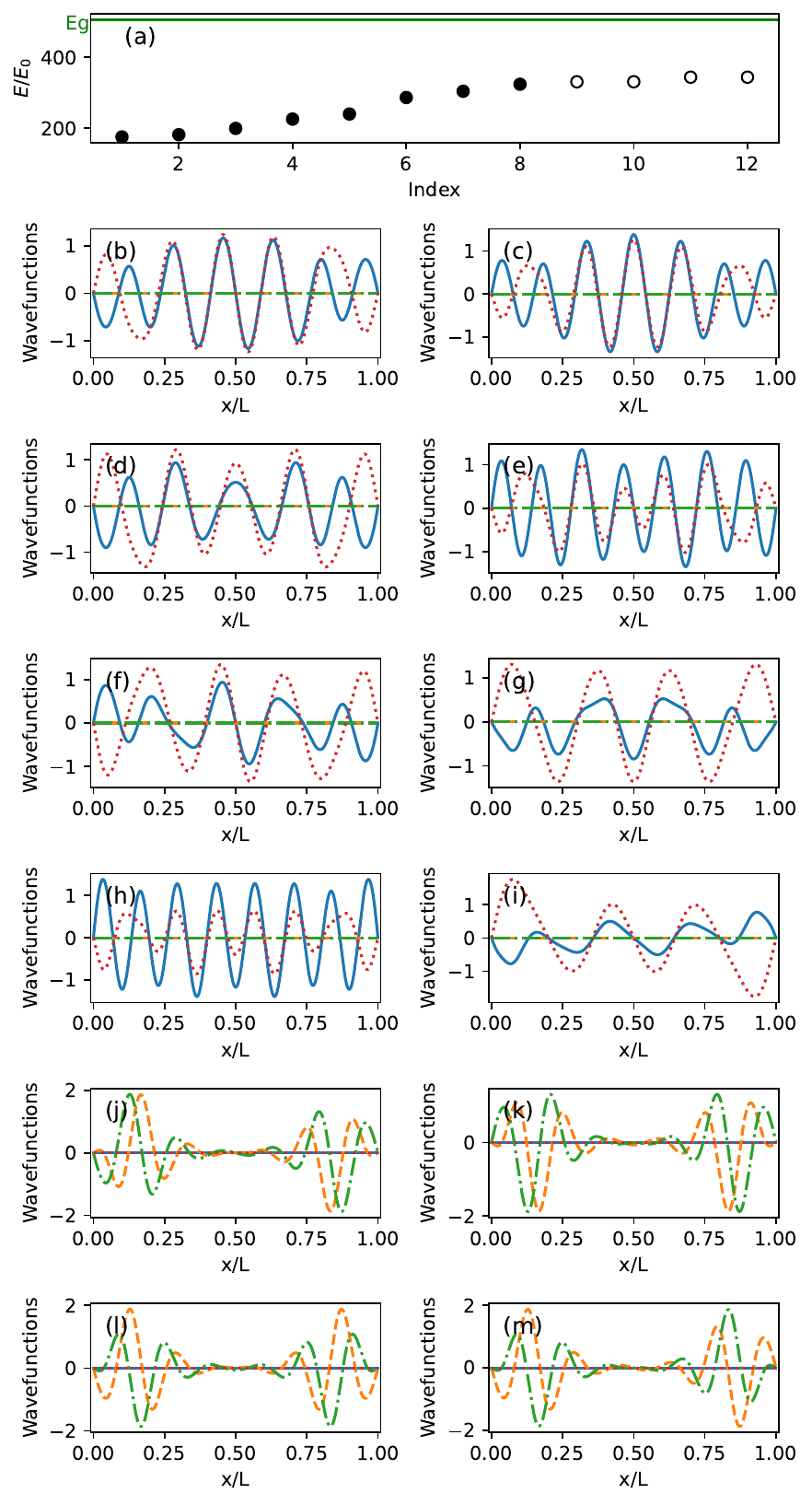}
\caption{(a) Energy spectrum and (b)-(m) wavefunctions of the energetically lowest 12 states of a 1D FFLO superfluid with $\mu_\downarrow=\mu_\uparrow-h$ on the BCS side. In panel (a), the solid (hollow) circles contribute to $n_\downarrow$ ($n_\uparrow$) while the solid line indicates the energy gap $E_g=\Delta$ determined by the bulk value.
Panels (b)-(m) illustrate the wavefunctions of the states in panel (a) from the lowest to the highest energies. The eigenfunctions $u_{\uparrow}$, $v_{\uparrow}$, $u_{\downarrow}$, $v_{\downarrow}$ are represented by the solid, dash, dot-dash, and dotted lines, respectively. Here, $U/(E_0 L)=-50$, $\mu_\uparrow/E_0=1000$, $h/E_0=675$, $N_\uparrow=11.5$, and $N_\downarrow=9.5$.}
\label{Fig:BS_Homo}
\end{figure}

The wavefunctions of the lowest twelve states lying below the energy gap $E_g=\Delta_0$ (determined by the bulk gap) which is marked by the plateau behavior of the gap function in Fig.~\ref{Fig:Profiles_Homo}(b) are presented in Fig.~\ref{Fig:BS_Homo}(b)-(m) with the associated energies depicted in panel (a). 
An important feature in the energy spectrum is that there are indeed low-energy states lying below the bulk gap energy $E_g$. Therefore, the FFLO superfluid introduces both a modulating order parameter and in-gap states. Among the lowest $12$ states, the lower-energy states contribute to $n_\downarrow$ while the higher ones contribute to $n_\uparrow$ because $\mu_\uparrow > \mu_\downarrow$. Therefore, the polarization field establishes effective population separation in energy space, as 
was the case of spin-dependent ULRM potentials.

Finally, one can see that the $u$ and $v$ wave functions in Fig.~\ref{Fig:BS_Homo} are out-of-phase among each other at both hard-wall boundaries. Their products thus contribute negatively to the pairing gap, which explains the sign change near both hard-wall boundaries. Given the similarity with the FFLO state from a homogeneous polarization field, the negative gap function and out-of-phase in-gap pairing states in the population-imbalance cases of the spin-dependent ULRM potentials may be attributed to the formation of a local FFLO-like state in the polarization region.


\begin{thebibliography}{71}%
	\makeatletter
	\providecommand \@ifxundefined [1]{%
		\@ifx{#1\undefined}
	}%
	\providecommand \@ifnum [1]{%
		\ifnum #1\expandafter \@firstoftwo
		\else \expandafter \@secondoftwo
		\fi
	}%
	\providecommand \@ifx [1]{%
		\ifx #1\expandafter \@firstoftwo
		\else \expandafter \@secondoftwo
		\fi
	}%
	\providecommand \natexlab [1]{#1}%
	\providecommand \enquote  [1]{``#1''}%
	\providecommand \bibnamefont  [1]{#1}%
	\providecommand \bibfnamefont [1]{#1}%
	\providecommand \citenamefont [1]{#1}%
	\providecommand \href@noop [0]{\@secondoftwo}%
	\providecommand \href [0]{\begingroup \@sanitize@url \@href}%
	\providecommand \@href[1]{\@@startlink{#1}\@@href}%
	\providecommand \@@href[1]{\endgroup#1\@@endlink}%
	\providecommand \@sanitize@url [0]{\catcode `\\12\catcode `\$12\catcode
		`\&12\catcode `\#12\catcode `\^12\catcode `\_12\catcode `\%12\relax}%
	\providecommand \@@startlink[1]{}%
	\providecommand \@@endlink[0]{}%
	\providecommand \url  [0]{\begingroup\@sanitize@url \@url }%
	\providecommand \@url [1]{\endgroup\@href {#1}{\urlprefix }}%
	\providecommand \urlprefix  [0]{URL }%
	\providecommand \Eprint [0]{\href }%
	\providecommand \doibase [0]{https://doi.org/}%
	\providecommand \selectlanguage [0]{\@gobble}%
	\providecommand \bibinfo  [0]{\@secondoftwo}%
	\providecommand \bibfield  [0]{\@secondoftwo}%
	\providecommand \translation [1]{[#1]}%
	\providecommand \BibitemOpen [0]{}%
	\providecommand \bibitemStop [0]{}%
	\providecommand \bibitemNoStop [0]{.\EOS\space}%
	\providecommand \EOS [0]{\spacefactor3000\relax}%
	\providecommand \BibitemShut  [1]{\csname bibitem#1\endcsname}%
	\let\auto@bib@innerbib\@empty
	\bibitem [{\citenamefont {Tinkham}(1996)}]{Tinkham_book}%
	\BibitemOpen
	\bibfield  {author} {\bibinfo {author} {\bibfnamefont {M.}~\bibnamefont
			{Tinkham}},\ }\href@noop {} {\emph {\bibinfo {title} {Introduction to
				Superconductivity}}},\ \bibinfo {edition} {2nd}\ ed.\ (\bibinfo  {publisher}
	{McGraw-Hill Book Co.},\ \bibinfo {address} {New York,NY},\ \bibinfo {year}
	{1996})\BibitemShut {NoStop}%
	\bibitem [{\citenamefont {Fetter}\ and\ \citenamefont
		{Walecka}(1971)}]{Fetter_book}%
	\BibitemOpen
	\bibfield  {author} {\bibinfo {author} {\bibfnamefont {A.~L.}\ \bibnamefont
			{Fetter}}\ and\ \bibinfo {author} {\bibfnamefont {J.~D.}\ \bibnamefont
			{Walecka}},\ }\href@noop {} {\emph {\bibinfo {title} {Quantum Theory of
				Many-Particle Systems}}}\ (\bibinfo  {publisher} {McGraw-Hill},\ \bibinfo
	{address} {Boston},\ \bibinfo {year} {1971})\BibitemShut {NoStop}%
	\bibitem [{\citenamefont {Chandrasekhar}(1962)}]{10.1063/1.1777362}%
	\BibitemOpen
	\bibfield  {author} {\bibinfo {author} {\bibfnamefont {B.~S.}\ \bibnamefont
			{Chandrasekhar}},\ }\bibfield  {title} {\bibinfo {title} {A note on the
			maximum critical field of high‐field superconductors},\ }\href
	{https://doi.org/10.1063/1.1777362} {\bibfield  {journal} {\bibinfo
			{journal} {Applied Physics Letters}\ }\textbf {\bibinfo {volume} {1}},\
		\bibinfo {pages} {7} (\bibinfo {year} {1962})},\ \Eprint
	{https://arxiv.org/abs/https://pubs.aip.org/aip/apl/article-pdf/1/1/7/18416028/7\_1\_online.pdf}
	{https://pubs.aip.org/aip/apl/article-pdf/1/1/7/18416028/7\_1\_online.pdf}
	\BibitemShut {NoStop}%
	\bibitem [{\citenamefont {Clogston}(1962)}]{PhysRevLett.9.266}%
	\BibitemOpen
	\bibfield  {author} {\bibinfo {author} {\bibfnamefont {A.~M.}\ \bibnamefont
			{Clogston}},\ }\bibfield  {title} {\bibinfo {title} {Upper limit for the
			critical field in hard superconductors},\ }\href
	{https://doi.org/10.1103/PhysRevLett.9.266} {\bibfield  {journal} {\bibinfo
			{journal} {Phys. Rev. Lett.}\ }\textbf {\bibinfo {volume} {9}},\ \bibinfo
		{pages} {266} (\bibinfo {year} {1962})}\BibitemShut {NoStop}%
	\bibitem [{\citenamefont {Fulde}\ and\ \citenamefont
		{Ferrell}(1964)}]{PhysRev.135.A550}%
	\BibitemOpen
	\bibfield  {author} {\bibinfo {author} {\bibfnamefont {P.}~\bibnamefont
			{Fulde}}\ and\ \bibinfo {author} {\bibfnamefont {R.~A.}\ \bibnamefont
			{Ferrell}},\ }\bibfield  {title} {\bibinfo {title} {Superconductivity in a
			strong spin-exchange field},\ }\href
	{https://doi.org/10.1103/PhysRev.135.A550} {\bibfield  {journal} {\bibinfo
			{journal} {Phys. Rev.}\ }\textbf {\bibinfo {volume} {135}},\ \bibinfo {pages}
		{A550} (\bibinfo {year} {1964})}\BibitemShut {NoStop}%
	\bibitem [{\citenamefont {Larkin}\ and\ \citenamefont
		{Ovchinnikov}(1964)}]{LO64}%
	\BibitemOpen
	\bibfield  {author} {\bibinfo {author} {\bibfnamefont {A.~I.}\ \bibnamefont
			{Larkin}}\ and\ \bibinfo {author} {\bibfnamefont {Y.~N.}\ \bibnamefont
			{Ovchinnikov}},\ }\bibfield  {title} {\bibinfo {title} {Nonuniform state of
			superconductors},\ }\bibfield  {journal} {\bibinfo  {journal} {Zh. Eksperim.
			Teor. Fiz.}\ }\textbf {\bibinfo {volume} {47}},\ \href
	{https://www.osti.gov/biblio/4653415} {} (\bibinfo {year} {1964})\BibitemShut
	{NoStop}%
	\bibitem [{\citenamefont {Dobrzyniecki}\ \emph {et~al.}(2021)\citenamefont
		{Dobrzyniecki}, \citenamefont {Orso},\ and\ \citenamefont
		{Sowi\ifmmode~\acute{n}\else \'{n}\fi{}ski}}]{PhysRevResearch.3.043105}%
	\BibitemOpen
	\bibfield  {author} {\bibinfo {author} {\bibfnamefont {J.}~\bibnamefont
			{Dobrzyniecki}}, \bibinfo {author} {\bibfnamefont {G.}~\bibnamefont {Orso}},\
		and\ \bibinfo {author} {\bibfnamefont {T.}~\bibnamefont
			{Sowi\ifmmode~\acute{n}\else \'{n}\fi{}ski}},\ }\bibfield  {title} {\bibinfo
		{title} {Unconventional pairing in few-fermion systems tuned by external
			confinement},\ }\href {https://doi.org/10.1103/PhysRevResearch.3.043105}
	{\bibfield  {journal} {\bibinfo  {journal} {Phys. Rev. Res.}\ }\textbf
		{\bibinfo {volume} {3}},\ \bibinfo {pages} {043105} (\bibinfo {year}
		{2021})}\BibitemShut {NoStop}%
	\bibitem [{\citenamefont {Steglich}\ \emph {et~al.}(1996)\citenamefont
		{Steglich}, \citenamefont {Modler}, \citenamefont {Gegenwart}, \citenamefont
		{Deppe}, \citenamefont {Weiden}, \citenamefont {Lang}, \citenamefont
		{Geibel}, \citenamefont {Lühmann}, \citenamefont {Paulsen}, \citenamefont
		{Tholence}, \citenamefont {Ōnuki}, \citenamefont {Tachiki},\ and\
		\citenamefont {Takahashi}}]{STEGLICH1996498}%
	\BibitemOpen
	\bibfield  {author} {\bibinfo {author} {\bibfnamefont {F.}~\bibnamefont
			{Steglich}}, \bibinfo {author} {\bibfnamefont {R.}~\bibnamefont {Modler}},
		\bibinfo {author} {\bibfnamefont {P.}~\bibnamefont {Gegenwart}}, \bibinfo
		{author} {\bibfnamefont {M.}~\bibnamefont {Deppe}}, \bibinfo {author}
		{\bibfnamefont {M.}~\bibnamefont {Weiden}}, \bibinfo {author} {\bibfnamefont
			{M.}~\bibnamefont {Lang}}, \bibinfo {author} {\bibfnamefont {C.}~\bibnamefont
			{Geibel}}, \bibinfo {author} {\bibfnamefont {T.}~\bibnamefont {Lühmann}},
		\bibinfo {author} {\bibfnamefont {C.}~\bibnamefont {Paulsen}}, \bibinfo
		{author} {\bibfnamefont {J.}~\bibnamefont {Tholence}}, \bibinfo {author}
		{\bibfnamefont {Y.}~\bibnamefont {Ōnuki}}, \bibinfo {author} {\bibfnamefont
			{M.}~\bibnamefont {Tachiki}},\ and\ \bibinfo {author} {\bibfnamefont
			{S.}~\bibnamefont {Takahashi}},\ }\bibfield  {title} {\bibinfo {title}
		{Experimental evidence for a generalized fflo state in clean type-ii
			superconductors with short coherence length and enhanced pauli
			susceptibility},\ }\href
	{https://doi.org/https://doi.org/10.1016/0921-4534(96)00065-2} {\bibfield
		{journal} {\bibinfo  {journal} {Physica C: Superconductivity}\ }\textbf
		{\bibinfo {volume} {263}},\ \bibinfo {pages} {498} (\bibinfo {year}
		{1996})},\ \bibinfo {note} {proceedings of the International Symposium on
		Frontiers of High - Tc Superconductivity}\BibitemShut {NoStop}%
	\bibitem [{\citenamefont {Imajo}\ \emph {et~al.}(2022)\citenamefont {Imajo},
		\citenamefont {Nomura}, \citenamefont {Kohama},\ and\ \citenamefont
		{Kindo}}]{Imajo2022}%
	\BibitemOpen
	\bibfield  {author} {\bibinfo {author} {\bibfnamefont {S.}~\bibnamefont
			{Imajo}}, \bibinfo {author} {\bibfnamefont {T.}~\bibnamefont {Nomura}},
		\bibinfo {author} {\bibfnamefont {Y.}~\bibnamefont {Kohama}},\ and\ \bibinfo
		{author} {\bibfnamefont {K.}~\bibnamefont {Kindo}},\ }\bibfield  {title}
	{\bibinfo {title} {Emergent anisotropy in the
			fulde--ferrell--larkin--ovchinnikov state},\ }\href
	{https://doi.org/10.1038/s41467-022-33354-1} {\bibfield  {journal} {\bibinfo
			{journal} {Nature Communications}\ }\textbf {\bibinfo {volume} {13}},\
		\bibinfo {pages} {5590} (\bibinfo {year} {2022})}\BibitemShut {NoStop}%
	\bibitem [{\citenamefont {Kinjo}\ \emph {et~al.}(2022)\citenamefont {Kinjo},
		\citenamefont {Manago}, \citenamefont {Kitagawa}, \citenamefont {Mao},
		\citenamefont {Yonezawa}, \citenamefont {Maeno},\ and\ \citenamefont
		{Ishida}}]{doi:10.1126/science.abb0332}%
	\BibitemOpen
	\bibfield  {author} {\bibinfo {author} {\bibfnamefont {K.}~\bibnamefont
			{Kinjo}}, \bibinfo {author} {\bibfnamefont {M.}~\bibnamefont {Manago}},
		\bibinfo {author} {\bibfnamefont {S.}~\bibnamefont {Kitagawa}}, \bibinfo
		{author} {\bibfnamefont {Z.~Q.}\ \bibnamefont {Mao}}, \bibinfo {author}
		{\bibfnamefont {S.}~\bibnamefont {Yonezawa}}, \bibinfo {author}
		{\bibfnamefont {Y.}~\bibnamefont {Maeno}},\ and\ \bibinfo {author}
		{\bibfnamefont {K.}~\bibnamefont {Ishida}},\ }\bibfield  {title} {\bibinfo
		{title} {Superconducting spin smecticity evidencing the
			fulde-ferrell-larkin-ovchinnikov state in sr$_2$ruo$_4$},\ }\href
	{https://doi.org/10.1126/science.abb0332} {\bibfield  {journal} {\bibinfo
			{journal} {Science}\ }\textbf {\bibinfo {volume} {376}},\ \bibinfo {pages}
		{397} (\bibinfo {year} {2022})},\ \Eprint
	{https://arxiv.org/abs/https://www.science.org/doi/pdf/10.1126/science.abb0332}
	{https://www.science.org/doi/pdf/10.1126/science.abb0332} \BibitemShut
	{NoStop}%
	\bibitem [{\citenamefont {Wan}\ \emph {et~al.}(2023)\citenamefont {Wan},
		\citenamefont {Zheliuk}, \citenamefont {Yuan}, \citenamefont {Peng},
		\citenamefont {Zhang}, \citenamefont {Liang}, \citenamefont {Zeitler},
		\citenamefont {Wiedmann}, \citenamefont {Hussey}, \citenamefont {Palstra},\
		and\ \citenamefont {Ye}}]{Wan2023}%
	\BibitemOpen
	\bibfield  {author} {\bibinfo {author} {\bibfnamefont {P.}~\bibnamefont
			{Wan}}, \bibinfo {author} {\bibfnamefont {O.}~\bibnamefont {Zheliuk}},
		\bibinfo {author} {\bibfnamefont {N.~F.~Q.}\ \bibnamefont {Yuan}}, \bibinfo
		{author} {\bibfnamefont {X.}~\bibnamefont {Peng}}, \bibinfo {author}
		{\bibfnamefont {L.}~\bibnamefont {Zhang}}, \bibinfo {author} {\bibfnamefont
			{M.}~\bibnamefont {Liang}}, \bibinfo {author} {\bibfnamefont
			{U.}~\bibnamefont {Zeitler}}, \bibinfo {author} {\bibfnamefont
			{S.}~\bibnamefont {Wiedmann}}, \bibinfo {author} {\bibfnamefont {N.~E.}\
			\bibnamefont {Hussey}}, \bibinfo {author} {\bibfnamefont {T.~T.~M.}\
			\bibnamefont {Palstra}},\ and\ \bibinfo {author} {\bibfnamefont
			{J.}~\bibnamefont {Ye}},\ }\bibfield  {title} {\bibinfo {title} {Orbital
			fulde--ferrell--larkin--ovchinnikov state in an ising superconductor},\
	}\href {https://doi.org/10.1038/s41586-023-05967-z} {\bibfield  {journal}
		{\bibinfo  {journal} {Nature}\ }\textbf {\bibinfo {volume} {619}},\ \bibinfo
		{pages} {46} (\bibinfo {year} {2023})}\BibitemShut {NoStop}%
	\bibitem [{\citenamefont {Liao}\ \emph {et~al.}(2010)\citenamefont {Liao},
		\citenamefont {Rittner}, \citenamefont {Paprotta}, \citenamefont {Li},
		\citenamefont {Partridge}, \citenamefont {Hulet}, \citenamefont {Baur},\ and\
		\citenamefont {Mueller}}]{Liao2010}%
	\BibitemOpen
	\bibfield  {author} {\bibinfo {author} {\bibfnamefont {Y.-a.}\ \bibnamefont
			{Liao}}, \bibinfo {author} {\bibfnamefont {A.~S.~C.}\ \bibnamefont
			{Rittner}}, \bibinfo {author} {\bibfnamefont {T.}~\bibnamefont {Paprotta}},
		\bibinfo {author} {\bibfnamefont {W.}~\bibnamefont {Li}}, \bibinfo {author}
		{\bibfnamefont {G.~B.}\ \bibnamefont {Partridge}}, \bibinfo {author}
		{\bibfnamefont {R.~G.}\ \bibnamefont {Hulet}}, \bibinfo {author}
		{\bibfnamefont {S.~K.}\ \bibnamefont {Baur}},\ and\ \bibinfo {author}
		{\bibfnamefont {E.~J.}\ \bibnamefont {Mueller}},\ }\bibfield  {title}
	{\bibinfo {title} {Spin-imbalance in a one-dimensional fermi gas},\ }\href
	{https://doi.org/10.1038/nature09393} {\bibfield  {journal} {\bibinfo
			{journal} {Nature}\ }\textbf {\bibinfo {volume} {467}},\ \bibinfo {pages}
		{567} (\bibinfo {year} {2010})}\BibitemShut {NoStop}%
	\bibitem [{\citenamefont {Revelle}\ \emph {et~al.}(2016)\citenamefont
		{Revelle}, \citenamefont {Fry}, \citenamefont {Olsen},\ and\ \citenamefont
		{Hulet}}]{PhysRevLett.117.235301}%
	\BibitemOpen
	\bibfield  {author} {\bibinfo {author} {\bibfnamefont {M.~C.}\ \bibnamefont
			{Revelle}}, \bibinfo {author} {\bibfnamefont {J.~A.}\ \bibnamefont {Fry}},
		\bibinfo {author} {\bibfnamefont {B.~A.}\ \bibnamefont {Olsen}},\ and\
		\bibinfo {author} {\bibfnamefont {R.~G.}\ \bibnamefont {Hulet}},\ }\bibfield
	{title} {\bibinfo {title} {1d to 3d {C}rossover of a {S}pin-{I}mbalanced
			{F}ermi {G}as},\ }\href {https://doi.org/10.1103/PhysRevLett.117.235301}
	{\bibfield  {journal} {\bibinfo  {journal} {Phys. Rev. Lett.}\ }\textbf
		{\bibinfo {volume} {117}},\ \bibinfo {pages} {235301} (\bibinfo {year}
		{2016})}\BibitemShut {NoStop}%
	\bibitem [{\citenamefont {Dupuis}(1995)}]{PhysRevB.51.9074}%
	\BibitemOpen
	\bibfield  {author} {\bibinfo {author} {\bibfnamefont {N.}~\bibnamefont
			{Dupuis}},\ }\bibfield  {title} {\bibinfo {title}
		{Larkin-ovchinnikov-fulde-ferrell state in quasi-one-dimensional
			superconductors},\ }\href {https://doi.org/10.1103/PhysRevB.51.9074}
	{\bibfield  {journal} {\bibinfo  {journal} {Phys. Rev. B}\ }\textbf {\bibinfo
			{volume} {51}},\ \bibinfo {pages} {9074} (\bibinfo {year}
		{1995})}\BibitemShut {NoStop}%
	\bibitem [{\citenamefont {Matsuda}\ and\ \citenamefont
		{Shimahara}(2007)}]{doi:10.1143/JPSJ.76.051005}%
	\BibitemOpen
	\bibfield  {author} {\bibinfo {author} {\bibfnamefont {Y.}~\bibnamefont
			{Matsuda}}\ and\ \bibinfo {author} {\bibfnamefont {H.}~\bibnamefont
			{Shimahara}},\ }\bibfield  {title} {\bibinfo {title}
		{Fulde–ferrell–larkin–ovchinnikov state in heavy fermion
			superconductors},\ }\href {https://doi.org/10.1143/JPSJ.76.051005} {\bibfield
		{journal} {\bibinfo  {journal} {Journal of the Physical Society of Japan}\
		}\textbf {\bibinfo {volume} {76}},\ \bibinfo {pages} {051005} (\bibinfo
		{year} {2007})},\ \Eprint
	{https://arxiv.org/abs/https://doi.org/10.1143/JPSJ.76.051005}
	{https://doi.org/10.1143/JPSJ.76.051005} \BibitemShut {NoStop}%
	\bibitem [{\citenamefont {He}\ \emph {et~al.}(2006)\citenamefont {He},
		\citenamefont {Jin},\ and\ \citenamefont {Zhuang}}]{PhysRevB.73.214527}%
	\BibitemOpen
	\bibfield  {author} {\bibinfo {author} {\bibfnamefont {L.}~\bibnamefont
			{He}}, \bibinfo {author} {\bibfnamefont {M.}~\bibnamefont {Jin}},\ and\
		\bibinfo {author} {\bibfnamefont {P.}~\bibnamefont {Zhuang}},\ }\bibfield
	{title} {\bibinfo {title} {Loff pairing vs breached pairing in asymmetric
			fermion superfluids},\ }\href {https://doi.org/10.1103/PhysRevB.73.214527}
	{\bibfield  {journal} {\bibinfo  {journal} {Phys. Rev. B}\ }\textbf {\bibinfo
			{volume} {73}},\ \bibinfo {pages} {214527} (\bibinfo {year}
		{2006})}\BibitemShut {NoStop}%
	\bibitem [{\citenamefont {Chen}\ \emph {et~al.}(2007)\citenamefont {Chen},
		\citenamefont {He}, \citenamefont {Chien},\ and\ \citenamefont
		{Levin}}]{PhysRevB.75.014521}%
	\BibitemOpen
	\bibfield  {author} {\bibinfo {author} {\bibfnamefont {Q.}~\bibnamefont
			{Chen}}, \bibinfo {author} {\bibfnamefont {Y.}~\bibnamefont {He}}, \bibinfo
		{author} {\bibfnamefont {C.-C.}\ \bibnamefont {Chien}},\ and\ \bibinfo
		{author} {\bibfnamefont {K.}~\bibnamefont {Levin}},\ }\bibfield  {title}
	{\bibinfo {title} {Theory of superfluids with population imbalance:
			Finite-temperature and bcs-bec crossover effects},\ }\href
	{https://doi.org/10.1103/PhysRevB.75.014521} {\bibfield  {journal} {\bibinfo
			{journal} {Phys. Rev. B}\ }\textbf {\bibinfo {volume} {75}},\ \bibinfo
		{pages} {014521} (\bibinfo {year} {2007})}\BibitemShut {NoStop}%
	\bibitem [{\citenamefont {Yu}(1965)}]{YU1965}%
	\BibitemOpen
	\bibfield  {author} {\bibinfo {author} {\bibfnamefont {L.}~\bibnamefont
			{Yu}},\ }\bibfield  {title} {\bibinfo {title} {Bound state in superconductors
			with paramagnetic impurities},\ }\href {https://doi.org/10.7498/aps.21.75}
	{\bibfield  {journal} {\bibinfo  {journal} {Acta Physica Sinica}\ }\textbf
		{\bibinfo {volume} {21}},\ \bibinfo {pages} {75} (\bibinfo {year}
		{1965})}\BibitemShut {NoStop}%
	\bibitem [{\citenamefont {Shiba}(1968)}]{Shiba68}%
	\BibitemOpen
	\bibfield  {author} {\bibinfo {author} {\bibfnamefont {H.}~\bibnamefont
			{Shiba}},\ }\bibfield  {title} {\bibinfo {title} {Classical spins in
			superconductors},\ }\href {https://doi.org/10.1143/PTP.40.435} {\bibfield
		{journal} {\bibinfo  {journal} {Progress of Theoretical Physics}\ }\textbf
		{\bibinfo {volume} {40}},\ \bibinfo {pages} {435} (\bibinfo {year} {1968})},\
	\Eprint
	{https://arxiv.org/abs/https://academic.oup.com/ptp/article-pdf/40/3/435/5185550/40-3-435.pdf}
	{https://academic.oup.com/ptp/article-pdf/40/3/435/5185550/40-3-435.pdf}
	\BibitemShut {NoStop}%
	\bibitem [{\citenamefont {Rusinov}(1969)}]{Rusinov69}%
	\BibitemOpen
	\bibfield  {author} {\bibinfo {author} {\bibfnamefont {A.~I.}\ \bibnamefont
			{Rusinov}},\ }\bibfield  {title} {\bibinfo {title} {Superconductivity near a
			paramagnetic impurity},\ }\href@noop {} {\bibfield  {journal} {\bibinfo
			{journal} {JETP Lett. (USSR)}\ }\textbf {\bibinfo {volume} {9}},\ \bibinfo
		{pages} {85} (\bibinfo {year} {1969})}\BibitemShut {NoStop}%
	\bibitem [{\citenamefont {Balatsky}\ \emph {et~al.}(2006)\citenamefont
		{Balatsky}, \citenamefont {Vekhter},\ and\ \citenamefont
		{Zhu}}]{RevModPhys.78.373}%
	\BibitemOpen
	\bibfield  {author} {\bibinfo {author} {\bibfnamefont {A.~V.}\ \bibnamefont
			{Balatsky}}, \bibinfo {author} {\bibfnamefont {I.}~\bibnamefont {Vekhter}},\
		and\ \bibinfo {author} {\bibfnamefont {J.-X.}\ \bibnamefont {Zhu}},\
	}\bibfield  {title} {\bibinfo {title} {Impurity-induced states in
			conventional and unconventional superconductors},\ }\href
	{https://doi.org/10.1103/RevModPhys.78.373} {\bibfield  {journal} {\bibinfo
			{journal} {Rev. Mod. Phys.}\ }\textbf {\bibinfo {volume} {78}},\ \bibinfo
		{pages} {373} (\bibinfo {year} {2006})}\BibitemShut {NoStop}%
	\bibitem [{\citenamefont {Heinrich}\ \emph {et~al.}(2018)\citenamefont
		{Heinrich}, \citenamefont {Pascual},\ and\ \citenamefont
		{Franke}}]{HEINRICH20181}%
	\BibitemOpen
	\bibfield  {author} {\bibinfo {author} {\bibfnamefont {B.~W.}\ \bibnamefont
			{Heinrich}}, \bibinfo {author} {\bibfnamefont {J.~I.}\ \bibnamefont
			{Pascual}},\ and\ \bibinfo {author} {\bibfnamefont {K.~J.}\ \bibnamefont
			{Franke}},\ }\bibfield  {title} {\bibinfo {title} {Single magnetic adsorbates
			on s-wave superconductors},\ }\href
	{https://doi.org/https://doi.org/10.1016/j.progsurf.2018.01.001} {\bibfield
		{journal} {\bibinfo  {journal} {Progress in Surface Science}\ }\textbf
		{\bibinfo {volume} {93}},\ \bibinfo {pages} {1} (\bibinfo {year}
		{2018})}\BibitemShut {NoStop}%
	\bibitem [{\citenamefont {Wang}\ \emph
		{et~al.}(2022{\natexlab{a}})\citenamefont {Wang}, \citenamefont {Liu},\ and\
		\citenamefont {Hu}}]{PhysRevLett.128.175301}%
	\BibitemOpen
	\bibfield  {author} {\bibinfo {author} {\bibfnamefont {J.}~\bibnamefont
			{Wang}}, \bibinfo {author} {\bibfnamefont {X.-J.}\ \bibnamefont {Liu}},\ and\
		\bibinfo {author} {\bibfnamefont {H.}~\bibnamefont {Hu}},\ }\bibfield
	{title} {\bibinfo {title} {Exact quasiparticle properties of a heavy polaron
			in bcs fermi superfluids},\ }\href
	{https://doi.org/10.1103/PhysRevLett.128.175301} {\bibfield  {journal}
		{\bibinfo  {journal} {Phys. Rev. Lett.}\ }\textbf {\bibinfo {volume} {128}},\
		\bibinfo {pages} {175301} (\bibinfo {year} {2022}{\natexlab{a}})}\BibitemShut
	{NoStop}%
	\bibitem [{\citenamefont {Wang}\ \emph
		{et~al.}(2022{\natexlab{b}})\citenamefont {Wang}, \citenamefont {Liu},\ and\
		\citenamefont {Hu}}]{PhysRevA.105.043320}%
	\BibitemOpen
	\bibfield  {author} {\bibinfo {author} {\bibfnamefont {J.}~\bibnamefont
			{Wang}}, \bibinfo {author} {\bibfnamefont {X.-J.}\ \bibnamefont {Liu}},\ and\
		\bibinfo {author} {\bibfnamefont {H.}~\bibnamefont {Hu}},\ }\bibfield
	{title} {\bibinfo {title} {Heavy polarons in ultracold atomic fermi
			superfluids at the bec-bcs crossover: Formalism and applications},\ }\href
	{https://doi.org/10.1103/PhysRevA.105.043320} {\bibfield  {journal} {\bibinfo
			{journal} {Phys. Rev. A}\ }\textbf {\bibinfo {volume} {105}},\ \bibinfo
		{pages} {043320} (\bibinfo {year} {2022}{\natexlab{b}})}\BibitemShut
	{NoStop}%
	\bibitem [{\citenamefont {Vernier}\ \emph {et~al.}(2011)\citenamefont
		{Vernier}, \citenamefont {Pekker}, \citenamefont {Zwierlein},\ and\
		\citenamefont {Demler}}]{PhysRevA.83.033619}%
	\BibitemOpen
	\bibfield  {author} {\bibinfo {author} {\bibfnamefont {E.}~\bibnamefont
			{Vernier}}, \bibinfo {author} {\bibfnamefont {D.}~\bibnamefont {Pekker}},
		\bibinfo {author} {\bibfnamefont {M.~W.}\ \bibnamefont {Zwierlein}},\ and\
		\bibinfo {author} {\bibfnamefont {E.}~\bibnamefont {Demler}},\ }\bibfield
	{title} {\bibinfo {title} {Bound states of a localized magnetic impurity in a
			superfluid of paired ultracold fermions},\ }\href
	{https://doi.org/10.1103/PhysRevA.83.033619} {\bibfield  {journal} {\bibinfo
			{journal} {Phys. Rev. A}\ }\textbf {\bibinfo {volume} {83}},\ \bibinfo
		{pages} {033619} (\bibinfo {year} {2011})}\BibitemShut {NoStop}%
	\bibitem [{\citenamefont {Greene}\ \emph {et~al.}(2000)\citenamefont {Greene},
		\citenamefont {Dickinson},\ and\ \citenamefont
		{Sadeghpour}}]{PhysRevLett.85.2458}%
	\BibitemOpen
	\bibfield  {author} {\bibinfo {author} {\bibfnamefont {C.~H.}\ \bibnamefont
			{Greene}}, \bibinfo {author} {\bibfnamefont {A.~S.}\ \bibnamefont
			{Dickinson}},\ and\ \bibinfo {author} {\bibfnamefont {H.~R.}\ \bibnamefont
			{Sadeghpour}},\ }\bibfield  {title} {\bibinfo {title} {Creation of {P}olar
			and {N}onpolar {U}ltra-{L}ong-{R}ange {R}ydberg {M}olecules},\ }\href
	{https://doi.org/10.1103/PhysRevLett.85.2458} {\bibfield  {journal} {\bibinfo
			{journal} {Phys. Rev. Lett.}\ }\textbf {\bibinfo {volume} {85}},\ \bibinfo
		{pages} {2458} (\bibinfo {year} {2000})}\BibitemShut {NoStop}%
	\bibitem [{\citenamefont {Chibisov}\ \emph {et~al.}(2002)\citenamefont
		{Chibisov}, \citenamefont {Khuskivadze},\ and\ \citenamefont
		{Fabrikant}}]{Chibisov02}%
	\BibitemOpen
	\bibfield  {author} {\bibinfo {author} {\bibfnamefont {M.~I.}\ \bibnamefont
			{Chibisov}}, \bibinfo {author} {\bibfnamefont {A.~A.}\ \bibnamefont
			{Khuskivadze}},\ and\ \bibinfo {author} {\bibfnamefont {I.~I.}\ \bibnamefont
			{Fabrikant}},\ }\bibfield  {title} {\bibinfo {title} {Energies and dipole
			moments of long-range molecular {R}ydberg states},\ }\href
	{https://doi.org/10.1088/0953-4075/35/10/101} {\bibfield  {journal} {\bibinfo
			{journal} {J. Phys. B: At. Mol. Opt. Phys.}\ }\textbf {\bibinfo {volume}
			{35}},\ \bibinfo {pages} {L193} (\bibinfo {year} {2002})}\BibitemShut
	{NoStop}%
	\bibitem [{\citenamefont {Hamilton}\ \emph {et~al.}(2002)\citenamefont
		{Hamilton}, \citenamefont {Greene},\ and\ \citenamefont
		{Sadeghpour}}]{Hamilton02}%
	\BibitemOpen
	\bibfield  {author} {\bibinfo {author} {\bibfnamefont {E.~L.}\ \bibnamefont
			{Hamilton}}, \bibinfo {author} {\bibfnamefont {C.~H.}\ \bibnamefont
			{Greene}},\ and\ \bibinfo {author} {\bibfnamefont {H.~R.}\ \bibnamefont
			{Sadeghpour}},\ }\bibfield  {title} {\bibinfo {title}
		{Shape-resonance-induced long-range molecular {R}ydberg states},\ }\href
	{https://doi.org/10.1088/0953-4075/35/10/102} {\bibfield  {journal} {\bibinfo
			{journal} {J. Phys. B: At. Mol. Opt. Phys.}\ }\textbf {\bibinfo {volume}
			{35}},\ \bibinfo {pages} {L199} (\bibinfo {year} {2002})}\BibitemShut
	{NoStop}%
	\bibitem [{\citenamefont {Bendkowsky}\ \emph {et~al.}(2009)\citenamefont
		{Bendkowsky}, \citenamefont {Butscher}, \citenamefont {Nipper}, \citenamefont
		{Shaffer}, \citenamefont {L{\"o}w},\ and\ \citenamefont
		{Pfau}}]{Bendkowsky2009}%
	\BibitemOpen
	\bibfield  {author} {\bibinfo {author} {\bibfnamefont {V.}~\bibnamefont
			{Bendkowsky}}, \bibinfo {author} {\bibfnamefont {B.}~\bibnamefont
			{Butscher}}, \bibinfo {author} {\bibfnamefont {J.}~\bibnamefont {Nipper}},
		\bibinfo {author} {\bibfnamefont {J.~P.}\ \bibnamefont {Shaffer}}, \bibinfo
		{author} {\bibfnamefont {R.}~\bibnamefont {L{\"o}w}},\ and\ \bibinfo {author}
		{\bibfnamefont {T.}~\bibnamefont {Pfau}},\ }\bibfield  {title} {\bibinfo
		{title} {Observation of ultralong-range rydberg molecules},\ }\href
	{https://doi.org/10.1038/nature07945} {\bibfield  {journal} {\bibinfo
			{journal} {Nature}\ }\textbf {\bibinfo {volume} {458}},\ \bibinfo {pages}
		{1005} (\bibinfo {year} {2009})}\BibitemShut {NoStop}%
	\bibitem [{\citenamefont {Niederpr{\"u}m}\ \emph {et~al.}(2016)\citenamefont
		{Niederpr{\"u}m}, \citenamefont {Thomas}, \citenamefont {Eichert},
		\citenamefont {Lippe}, \citenamefont {P{\'e}rez-R{\'i}os}, \citenamefont
		{Greene},\ and\ \citenamefont {Ott}}]{Niederprum2016}%
	\BibitemOpen
	\bibfield  {author} {\bibinfo {author} {\bibfnamefont {T.}~\bibnamefont
			{Niederpr{\"u}m}}, \bibinfo {author} {\bibfnamefont {O.}~\bibnamefont
			{Thomas}}, \bibinfo {author} {\bibfnamefont {T.}~\bibnamefont {Eichert}},
		\bibinfo {author} {\bibfnamefont {C.}~\bibnamefont {Lippe}}, \bibinfo
		{author} {\bibfnamefont {J.}~\bibnamefont {P{\'e}rez-R{\'i}os}}, \bibinfo
		{author} {\bibfnamefont {C.~H.}\ \bibnamefont {Greene}},\ and\ \bibinfo
		{author} {\bibfnamefont {H.}~\bibnamefont {Ott}},\ }\bibfield  {title}
	{\bibinfo {title} {Observation of pendular butterfly {R}ydberg molecules},\
	}\href {https://doi.org/10.1038/ncomms12820} {\bibfield  {journal} {\bibinfo
			{journal} {Nature Communications}\ }\textbf {\bibinfo {volume} {7}},\
		\bibinfo {pages} {12820} (\bibinfo {year} {2016})}\BibitemShut {NoStop}%
	\bibitem [{\citenamefont {Alth{\"o}n}\ \emph {et~al.}(2023)\citenamefont
		{Alth{\"o}n}, \citenamefont {Exner}, \citenamefont {Bl{\"a}ttner},\ and\
		\citenamefont {Ott}}]{Althon2023}%
	\BibitemOpen
	\bibfield  {author} {\bibinfo {author} {\bibfnamefont {M.}~\bibnamefont
			{Alth{\"o}n}}, \bibinfo {author} {\bibfnamefont {M.}~\bibnamefont {Exner}},
		\bibinfo {author} {\bibfnamefont {R.}~\bibnamefont {Bl{\"a}ttner}},\ and\
		\bibinfo {author} {\bibfnamefont {H.}~\bibnamefont {Ott}},\ }\bibfield
	{title} {\bibinfo {title} {Exploring the vibrational series of pure trilobite
			{R}ydberg molecules},\ }\href {https://doi.org/10.1038/s41467-023-43818-7}
	{\bibfield  {journal} {\bibinfo  {journal} {Nature Communications}\ }\textbf
		{\bibinfo {volume} {14}},\ \bibinfo {pages} {8108} (\bibinfo {year}
		{2023})}\BibitemShut {NoStop}%
	\bibitem [{\citenamefont {Booth}\ \emph {et~al.}(2015)\citenamefont {Booth},
		\citenamefont {Rittenhouse}, \citenamefont {Yang}, \citenamefont
		{Sadeghpour},\ and\ \citenamefont {Shaffer}}]{Booth15}%
	\BibitemOpen
	\bibfield  {author} {\bibinfo {author} {\bibfnamefont {D.}~\bibnamefont
			{Booth}}, \bibinfo {author} {\bibfnamefont {S.~T.}\ \bibnamefont
			{Rittenhouse}}, \bibinfo {author} {\bibfnamefont {J.}~\bibnamefont {Yang}},
		\bibinfo {author} {\bibfnamefont {H.~R.}\ \bibnamefont {Sadeghpour}},\ and\
		\bibinfo {author} {\bibfnamefont {J.~P.}\ \bibnamefont {Shaffer}},\
	}\bibfield  {title} {\bibinfo {title} {Production of trilobite {R}ydberg
			molecule dimers with kilo-{D}ebye permanent electric dipole moments},\ }\href
	{https://doi.org/10.1126/science.1260722} {\bibfield  {journal} {\bibinfo
			{journal} {Science}\ }\textbf {\bibinfo {volume} {348}},\ \bibinfo {pages}
		{99} (\bibinfo {year} {2015})},\ \Eprint
	{https://arxiv.org/abs/https://www.science.org/doi/pdf/10.1126/science.1260722}
	{https://www.science.org/doi/pdf/10.1126/science.1260722} \BibitemShut
	{NoStop}%
	\bibitem [{\citenamefont {Whalen}\ \emph {et~al.}(2020)\citenamefont {Whalen},
		\citenamefont {Kanungo}, \citenamefont {Lu}, \citenamefont {Yoshida},
		\citenamefont {Burgd\"orfer}, \citenamefont {Dunning},\ and\ \citenamefont
		{Killian}}]{PhysRevA.101.060701}%
	\BibitemOpen
	\bibfield  {author} {\bibinfo {author} {\bibfnamefont {J.~D.}\ \bibnamefont
			{Whalen}}, \bibinfo {author} {\bibfnamefont {S.~K.}\ \bibnamefont {Kanungo}},
		\bibinfo {author} {\bibfnamefont {Y.}~\bibnamefont {Lu}}, \bibinfo {author}
		{\bibfnamefont {S.}~\bibnamefont {Yoshida}}, \bibinfo {author} {\bibfnamefont
			{J.}~\bibnamefont {Burgd\"orfer}}, \bibinfo {author} {\bibfnamefont {F.~B.}\
			\bibnamefont {Dunning}},\ and\ \bibinfo {author} {\bibfnamefont {T.~C.}\
			\bibnamefont {Killian}},\ }\bibfield  {title} {\bibinfo {title}
		{Heteronuclear rydberg molecules},\ }\href
	{https://doi.org/10.1103/PhysRevA.101.060701} {\bibfield  {journal} {\bibinfo
			{journal} {Phys. Rev. A}\ }\textbf {\bibinfo {volume} {101}},\ \bibinfo
		{pages} {060701} (\bibinfo {year} {2020})}\BibitemShut {NoStop}%
	\bibitem [{\citenamefont {Shaffer}\ \emph {et~al.}(2018)\citenamefont
		{Shaffer}, \citenamefont {Rittenhouse},\ and\ \citenamefont
		{Sadeghpour}}]{Shaffer2018}%
	\BibitemOpen
	\bibfield  {author} {\bibinfo {author} {\bibfnamefont {J.~P.}\ \bibnamefont
			{Shaffer}}, \bibinfo {author} {\bibfnamefont {S.~T.}\ \bibnamefont
			{Rittenhouse}},\ and\ \bibinfo {author} {\bibfnamefont {H.~R.}\ \bibnamefont
			{Sadeghpour}},\ }\bibfield  {title} {\bibinfo {title} {Ultracold {R}ydberg
			molecules},\ }\href {https://doi.org/10.1038/s41467-018-04135-6} {\bibfield
		{journal} {\bibinfo  {journal} {Nature Communications}\ }\textbf {\bibinfo
			{volume} {9}},\ \bibinfo {pages} {1965} (\bibinfo {year} {2018})}\BibitemShut
	{NoStop}%
	\bibitem [{\citenamefont {Christian~Fey}\ and\ \citenamefont
		{Schmelcher}(2020)}]{Fey20}%
	\BibitemOpen
	\bibfield  {author} {\bibinfo {author} {\bibfnamefont {F.~H.}\ \bibnamefont
			{Christian~Fey}}\ and\ \bibinfo {author} {\bibfnamefont {P.}~\bibnamefont
			{Schmelcher}},\ }\bibfield  {title} {\bibinfo {title} {Ultralong-range
			{R}ydberg molecules},\ }\href {https://doi.org/10.1080/00268976.2019.1679401}
	{\bibfield  {journal} {\bibinfo  {journal} {Molecular Physics}\ }\textbf
		{\bibinfo {volume} {118}},\ \bibinfo {pages} {e1679401} (\bibinfo {year}
		{2020})}\BibitemShut {NoStop}%
	\bibitem [{\citenamefont {Camargo}\ \emph {et~al.}(2018)\citenamefont
		{Camargo}, \citenamefont {Schmidt}, \citenamefont {Whalen}, \citenamefont
		{Ding}, \citenamefont {Woehl}, \citenamefont {Yoshida}, \citenamefont
		{Burgd\"orfer}, \citenamefont {Dunning}, \citenamefont {Sadeghpour},
		\citenamefont {Demler},\ and\ \citenamefont
		{Killian}}]{PhysRevLett.120.083401}%
	\BibitemOpen
	\bibfield  {author} {\bibinfo {author} {\bibfnamefont {F.}~\bibnamefont
			{Camargo}}, \bibinfo {author} {\bibfnamefont {R.}~\bibnamefont {Schmidt}},
		\bibinfo {author} {\bibfnamefont {J.~D.}\ \bibnamefont {Whalen}}, \bibinfo
		{author} {\bibfnamefont {R.}~\bibnamefont {Ding}}, \bibinfo {author}
		{\bibfnamefont {G.}~\bibnamefont {Woehl}}, \bibinfo {author} {\bibfnamefont
			{S.}~\bibnamefont {Yoshida}}, \bibinfo {author} {\bibfnamefont
			{J.}~\bibnamefont {Burgd\"orfer}}, \bibinfo {author} {\bibfnamefont {F.~B.}\
			\bibnamefont {Dunning}}, \bibinfo {author} {\bibfnamefont {H.~R.}\
			\bibnamefont {Sadeghpour}}, \bibinfo {author} {\bibfnamefont
			{E.}~\bibnamefont {Demler}},\ and\ \bibinfo {author} {\bibfnamefont {T.~C.}\
			\bibnamefont {Killian}},\ }\bibfield  {title} {\bibinfo {title} {Creation of
			{R}ydberg {P}olarons in a {B}ose {G}as},\ }\href
	{https://doi.org/10.1103/PhysRevLett.120.083401} {\bibfield  {journal}
		{\bibinfo  {journal} {Phys. Rev. Lett.}\ }\textbf {\bibinfo {volume} {120}},\
		\bibinfo {pages} {083401} (\bibinfo {year} {2018})}\BibitemShut {NoStop}%
	\bibitem [{\citenamefont {Schmidt}\ \emph {et~al.}(2016)\citenamefont
		{Schmidt}, \citenamefont {Sadeghpour},\ and\ \citenamefont
		{Demler}}]{PhysRevLett.116.105302}%
	\BibitemOpen
	\bibfield  {author} {\bibinfo {author} {\bibfnamefont {R.}~\bibnamefont
			{Schmidt}}, \bibinfo {author} {\bibfnamefont {H.~R.}\ \bibnamefont
			{Sadeghpour}},\ and\ \bibinfo {author} {\bibfnamefont {E.}~\bibnamefont
			{Demler}},\ }\bibfield  {title} {\bibinfo {title} {Mesoscopic {R}ydberg
			{I}mpurity in an {A}tomic {Q}uantum {G}as},\ }\href
	{https://doi.org/10.1103/PhysRevLett.116.105302} {\bibfield  {journal}
		{\bibinfo  {journal} {Phys. Rev. Lett.}\ }\textbf {\bibinfo {volume} {116}},\
		\bibinfo {pages} {105302} (\bibinfo {year} {2016})}\BibitemShut {NoStop}%
	\bibitem [{\citenamefont {Schmidt}\ \emph {et~al.}(2018)\citenamefont
		{Schmidt}, \citenamefont {Whalen}, \citenamefont {Ding}, \citenamefont
		{Camargo}, \citenamefont {Woehl}, \citenamefont {Yoshida}, \citenamefont
		{Burgd\"orfer}, \citenamefont {Dunning}, \citenamefont {Demler},
		\citenamefont {Sadeghpour},\ and\ \citenamefont
		{Killian}}]{PhysRevA.97.022707}%
	\BibitemOpen
	\bibfield  {author} {\bibinfo {author} {\bibfnamefont {R.}~\bibnamefont
			{Schmidt}}, \bibinfo {author} {\bibfnamefont {J.~D.}\ \bibnamefont {Whalen}},
		\bibinfo {author} {\bibfnamefont {R.}~\bibnamefont {Ding}}, \bibinfo {author}
		{\bibfnamefont {F.}~\bibnamefont {Camargo}}, \bibinfo {author} {\bibfnamefont
			{G.}~\bibnamefont {Woehl}}, \bibinfo {author} {\bibfnamefont
			{S.}~\bibnamefont {Yoshida}}, \bibinfo {author} {\bibfnamefont
			{J.}~\bibnamefont {Burgd\"orfer}}, \bibinfo {author} {\bibfnamefont {F.~B.}\
			\bibnamefont {Dunning}}, \bibinfo {author} {\bibfnamefont {E.}~\bibnamefont
			{Demler}}, \bibinfo {author} {\bibfnamefont {H.~R.}\ \bibnamefont
			{Sadeghpour}},\ and\ \bibinfo {author} {\bibfnamefont {T.~C.}\ \bibnamefont
			{Killian}},\ }\bibfield  {title} {\bibinfo {title} {Theory of excitation of
			{R}ydberg polarons in an atomic quantum gas},\ }\href
	{https://doi.org/10.1103/PhysRevA.97.022707} {\bibfield  {journal} {\bibinfo
			{journal} {Phys. Rev. A}\ }\textbf {\bibinfo {volume} {97}},\ \bibinfo
		{pages} {022707} (\bibinfo {year} {2018})}\BibitemShut {NoStop}%
	\bibitem [{\citenamefont {Sous}\ \emph {et~al.}(2020)\citenamefont {Sous},
		\citenamefont {Sadeghpour}, \citenamefont {Killian}, \citenamefont {Demler},\
		and\ \citenamefont {Schmidt}}]{Sous20}%
	\BibitemOpen
	\bibfield  {author} {\bibinfo {author} {\bibfnamefont {J.}~\bibnamefont
			{Sous}}, \bibinfo {author} {\bibfnamefont {H.~R.}\ \bibnamefont
			{Sadeghpour}}, \bibinfo {author} {\bibfnamefont {T.~C.}\ \bibnamefont
			{Killian}}, \bibinfo {author} {\bibfnamefont {E.}~\bibnamefont {Demler}},\
		and\ \bibinfo {author} {\bibfnamefont {R.}~\bibnamefont {Schmidt}},\
	}\bibfield  {title} {\bibinfo {title} {Rydberg impurity in a {F}ermi gas:
			{Q}uantum statistics and rotational blockade},\ }\href
	{https://doi.org/10.1103/PhysRevResearch.2.023021} {\bibfield  {journal}
		{\bibinfo  {journal} {Phys. Rev. Res.}\ }\textbf {\bibinfo {volume} {2}},\
		\bibinfo {pages} {023021} (\bibinfo {year} {2020})}\BibitemShut {NoStop}%
	\bibitem [{\citenamefont {Durst}\ and\ \citenamefont {Eiles}(2024)}]{Durst24}%
	\BibitemOpen
	\bibfield  {author} {\bibinfo {author} {\bibfnamefont {A.~A.~T.}\
			\bibnamefont {Durst}}\ and\ \bibinfo {author} {\bibfnamefont {M.~T.}\
			\bibnamefont {Eiles}},\ }\href@noop {} {\bibinfo {title} {Phenomenology of a
			{R}ydberg impurity in an ideal {B}ose {E}instein condensate}} (\bibinfo
	{year} {2024}),\ \bibinfo {note} {arXiv: 2404.03980}\BibitemShut {NoStop}%
	\bibitem [{\citenamefont {Chien}\ \emph {et~al.}(2024)\citenamefont {Chien},
		\citenamefont {Mistakidis},\ and\ \citenamefont {Sadeghpour}}]{RM_PRA24}%
	\BibitemOpen
	\bibfield  {author} {\bibinfo {author} {\bibfnamefont {C.-C.}\ \bibnamefont
			{Chien}}, \bibinfo {author} {\bibfnamefont {S.~I.}\ \bibnamefont
			{Mistakidis}},\ and\ \bibinfo {author} {\bibfnamefont {H.~R.}\ \bibnamefont
			{Sadeghpour}},\ }\bibfield  {title} {\bibinfo {title} {Breaking and trapping
			cooper pairs by rydberg-molecule spectroscopy in atomic fermi superfluids},\
	}\href {https://doi.org/10.1103/PhysRevA.110.L051303} {\bibfield  {journal}
		{\bibinfo  {journal} {Phys. Rev. A}\ }\textbf {\bibinfo {volume} {110}},\
		\bibinfo {pages} {L051303} (\bibinfo {year} {2024})}\BibitemShut {NoStop}%
	\bibitem [{\citenamefont {Eiles}\ and\ \citenamefont
		{Greene}(2017)}]{PhysRevA.95.042515}%
	\BibitemOpen
	\bibfield  {author} {\bibinfo {author} {\bibfnamefont {M.~T.}\ \bibnamefont
			{Eiles}}\ and\ \bibinfo {author} {\bibfnamefont {C.~H.}\ \bibnamefont
			{Greene}},\ }\bibfield  {title} {\bibinfo {title} {Hamiltonian for the
			inclusion of spin effects in long-range rydberg molecules},\ }\href
	{https://doi.org/10.1103/PhysRevA.95.042515} {\bibfield  {journal} {\bibinfo
			{journal} {Phys. Rev. A}\ }\textbf {\bibinfo {volume} {95}},\ \bibinfo
		{pages} {042515} (\bibinfo {year} {2017})}\BibitemShut {NoStop}%
	\bibitem [{\citenamefont {Eiles}(2018)}]{PhysRevA.98.042706}%
	\BibitemOpen
	\bibfield  {author} {\bibinfo {author} {\bibfnamefont {M.~T.}\ \bibnamefont
			{Eiles}},\ }\bibfield  {title} {\bibinfo {title} {Formation of long-range
			rydberg molecules in two-component ultracold gases},\ }\href
	{https://doi.org/10.1103/PhysRevA.98.042706} {\bibfield  {journal} {\bibinfo
			{journal} {Phys. Rev. A}\ }\textbf {\bibinfo {volume} {98}},\ \bibinfo
		{pages} {042706} (\bibinfo {year} {2018})}\BibitemShut {NoStop}%
	\bibitem [{\citenamefont {De~Gennes}(2018)}]{degennes-sc}%
	\BibitemOpen
	\bibfield  {author} {\bibinfo {author} {\bibfnamefont {P.~G.}\ \bibnamefont
			{De~Gennes}},\ }\href@noop {} {\emph {\bibinfo {title} {Superconductivity of
				Metals and Alloys.}}},\ \bibinfo {edition} {2nd}\ ed.,\ Advanced Books
	Classics\ (\bibinfo  {publisher} {Chapman and Hall/CRC},\ \bibinfo {address}
	{Boulder},\ \bibinfo {year} {2018})\BibitemShut {NoStop}%
	\bibitem [{\citenamefont {Zhu}(2016)}]{BdG-book}%
	\BibitemOpen
	\bibfield  {author} {\bibinfo {author} {\bibfnamefont {J.-X.}\ \bibnamefont
			{Zhu}},\ }\href@noop {} {\emph {\bibinfo {title} {Bogoliubov-de {G}ennes
				{M}ethod and {I}ts {A}pplications}}},\ \bibinfo {edition} {1st}\ ed.,\
	Lecture {N}otes in {P}hysics, 924\ (\bibinfo  {publisher} {Springer
		International Publishing},\ \bibinfo {address} {Cham, Switzerland},\ \bibinfo
	{year} {2016})\BibitemShut {NoStop}%
	\bibitem [{\citenamefont {Boche\'{n}ski}\ \emph {et~al.}(2024)\citenamefont
		{Boche\'{n}ski}, \citenamefont {Dobosz},\ and\ \citenamefont
		{Semczuk}}]{Bochenski24}%
	\BibitemOpen
	\bibfield  {author} {\bibinfo {author} {\bibfnamefont {M.}~\bibnamefont
			{Boche\'{n}ski}}, \bibinfo {author} {\bibfnamefont {J.}~\bibnamefont
			{Dobosz}},\ and\ \bibinfo {author} {\bibfnamefont {M.}~\bibnamefont
			{Semczuk}},\ }\bibfield  {title} {\bibinfo {title} {Magnetic trapping of an
			ultracold 39k-40k mixture with a versatile potassium laser system},\ }\href
	{https://doi.org/10.1364/OE.529071} {\bibfield  {journal} {\bibinfo
			{journal} {Opt. Express}\ }\textbf {\bibinfo {volume} {32}},\ \bibinfo
		{pages} {48463} (\bibinfo {year} {2024})}\BibitemShut {NoStop}%
	\bibitem [{\citenamefont {Wu}\ \emph {et~al.}(2011)\citenamefont {Wu},
		\citenamefont {Santiago}, \citenamefont {Park}, \citenamefont {Ahmadi},\ and\
		\citenamefont {Zwierlein}}]{PhysRevA.84.011601}%
	\BibitemOpen
	\bibfield  {author} {\bibinfo {author} {\bibfnamefont {C.-H.}\ \bibnamefont
			{Wu}}, \bibinfo {author} {\bibfnamefont {I.}~\bibnamefont {Santiago}},
		\bibinfo {author} {\bibfnamefont {J.~W.}\ \bibnamefont {Park}}, \bibinfo
		{author} {\bibfnamefont {P.}~\bibnamefont {Ahmadi}},\ and\ \bibinfo {author}
		{\bibfnamefont {M.~W.}\ \bibnamefont {Zwierlein}},\ }\bibfield  {title}
	{\bibinfo {title} {Strongly interacting isotopic bose-fermi mixture immersed
			in a fermi sea},\ }\href {https://doi.org/10.1103/PhysRevA.84.011601}
	{\bibfield  {journal} {\bibinfo  {journal} {Phys. Rev. A}\ }\textbf {\bibinfo
			{volume} {84}},\ \bibinfo {pages} {011601} (\bibinfo {year}
		{2011})}\BibitemShut {NoStop}%
	\bibitem [{\citenamefont {Touchard}\ \emph {et~al.}(1982)\citenamefont
		{Touchard}, \citenamefont {Guimbal}, \citenamefont {Büttgenbach},
		\citenamefont {Klapisch}, \citenamefont {{De Saint Simon}}, \citenamefont
		{Serre}, \citenamefont {Thibault}, \citenamefont {Duong}, \citenamefont
		{Juncar}, \citenamefont {Liberman}, \citenamefont {Pinard},\ and\
		\citenamefont {Vialle}}]{touchard1982}%
	\BibitemOpen
	\bibfield  {author} {\bibinfo {author} {\bibfnamefont {F.}~\bibnamefont
			{Touchard}}, \bibinfo {author} {\bibfnamefont {P.}~\bibnamefont {Guimbal}},
		\bibinfo {author} {\bibfnamefont {S.}~\bibnamefont {Büttgenbach}}, \bibinfo
		{author} {\bibfnamefont {R.}~\bibnamefont {Klapisch}}, \bibinfo {author}
		{\bibfnamefont {M.}~\bibnamefont {{De Saint Simon}}}, \bibinfo {author}
		{\bibfnamefont {J.}~\bibnamefont {Serre}}, \bibinfo {author} {\bibfnamefont
			{C.}~\bibnamefont {Thibault}}, \bibinfo {author} {\bibfnamefont
			{H.}~\bibnamefont {Duong}}, \bibinfo {author} {\bibfnamefont
			{P.}~\bibnamefont {Juncar}}, \bibinfo {author} {\bibfnamefont
			{S.}~\bibnamefont {Liberman}}, \bibinfo {author} {\bibfnamefont
			{J.}~\bibnamefont {Pinard}},\ and\ \bibinfo {author} {\bibfnamefont
			{J.}~\bibnamefont {Vialle}},\ }\bibfield  {title} {\bibinfo {title} {Isotope
			shifts and hyperfine structure of 38–47k by laser spectroscopy},\ }\href
	{https://doi.org/https://doi.org/10.1016/0370-2693(82)91167-4} {\bibfield
		{journal} {\bibinfo  {journal} {Physics Letters B}\ }\textbf {\bibinfo
			{volume} {108}},\ \bibinfo {pages} {169} (\bibinfo {year}
		{1982})}\BibitemShut {NoStop}%
	\bibitem [{\citenamefont {Anderson}\ \emph {et~al.}(2014)\citenamefont
		{Anderson}, \citenamefont {Miller},\ and\ \citenamefont
		{Raithel}}]{anderson2014}%
	\BibitemOpen
	\bibfield  {author} {\bibinfo {author} {\bibfnamefont {D.~A.}\ \bibnamefont
			{Anderson}}, \bibinfo {author} {\bibfnamefont {S.~A.}\ \bibnamefont
			{Miller}},\ and\ \bibinfo {author} {\bibfnamefont {G.}~\bibnamefont
			{Raithel}},\ }\bibfield  {title} {\bibinfo {title} {Angular-momentum
			couplings in long-range $\text{Rb}_2$ \text{Rydberg} molecules},\ }\href
	{https://doi.org/10.1103/PhysRevA.90.062518} {\bibfield  {journal} {\bibinfo
			{journal} {Phys. Rev. A}\ }\textbf {\bibinfo {volume} {90}},\ \bibinfo
		{pages} {062518} (\bibinfo {year} {2014})}\BibitemShut {NoStop}%
	\bibitem [{\citenamefont {Eiles}(2019)}]{eiles2019trilobites}%
	\BibitemOpen
	\bibfield  {author} {\bibinfo {author} {\bibfnamefont {M.~T.}\ \bibnamefont
			{Eiles}},\ }\bibfield  {title} {\bibinfo {title} {Trilobites, butterflies,
			and other exotic specimens of long-range rydberg molecules},\ }\href@noop {}
	{\bibfield  {journal} {\bibinfo  {journal} {Journal of Physics B: Atomic,
				Molecular and Optical Physics}\ }\textbf {\bibinfo {volume} {52}},\ \bibinfo
		{pages} {113001} (\bibinfo {year} {2019})}\BibitemShut {NoStop}%
	\bibitem [{\citenamefont {Karule}(1965)}]{karule1965elastic}%
	\BibitemOpen
	\bibfield  {author} {\bibinfo {author} {\bibfnamefont {E.}~\bibnamefont
			{Karule}},\ }\bibfield  {title} {\bibinfo {title} {Elastic scattering of
			low-energy electrons by alkali atoms},\ }\href@noop {} {\bibfield  {journal}
		{\bibinfo  {journal} {Physics Letters}\ }\textbf {\bibinfo {volume} {15}},\
		\bibinfo {pages} {137} (\bibinfo {year} {1965})}\BibitemShut {NoStop}%
	\bibitem [{\citenamefont {Lorenzen}\ and\ \citenamefont
		{Niemax}(1983)}]{lorenzen1983quantum}%
	\BibitemOpen
	\bibfield  {author} {\bibinfo {author} {\bibfnamefont {C.~J.}\ \bibnamefont
			{Lorenzen}}\ and\ \bibinfo {author} {\bibfnamefont {K.}~\bibnamefont
			{Niemax}},\ }\bibfield  {title} {\bibinfo {title} {Quantum defects of the
			n2p1/2, 3/2 levels in 39k i and 85rb i},\ }\href@noop {} {\bibfield
		{journal} {\bibinfo  {journal} {Physica Scripta}\ }\textbf {\bibinfo {volume}
			{27}},\ \bibinfo {pages} {300} (\bibinfo {year} {1983})}\BibitemShut
	{NoStop}%
	\bibitem [{\citenamefont {Leggett}(2006)}]{Leggett}%
	\BibitemOpen
	\bibfield  {author} {\bibinfo {author} {\bibfnamefont {A.~J.}\ \bibnamefont
			{Leggett}},\ }\href@noop {} {\emph {\bibinfo {title} {Quantum Liquids : Bose
				condensation and Cooper pairing in condensed-matter systems}}},\ Oxford
	Graduate Texts\ (\bibinfo  {publisher} {Oxford University Press},\ \bibinfo
	{address} {Oxford, UK},\ \bibinfo {year} {2006})\BibitemShut {NoStop}%
	\bibitem [{\citenamefont {Pethick}\ and\ \citenamefont
		{Smith}(2008)}]{Pethick-BEC}%
	\BibitemOpen
	\bibfield  {author} {\bibinfo {author} {\bibfnamefont {C.~J.}\ \bibnamefont
			{Pethick}}\ and\ \bibinfo {author} {\bibfnamefont {H.}~\bibnamefont
			{Smith}},\ }\href {https://doi.org/10.1017/CBO9780511802850} {\emph {\bibinfo
			{title} {Bose–{E}instein {C}ondensation in {D}ilute {G}ases}}},\ \bibinfo
	{edition} {2nd}\ ed.\ (\bibinfo  {publisher} {Cambridge University Press},\
	\bibinfo {year} {2008})\BibitemShut {NoStop}%
	\bibitem [{\citenamefont {Olshanii}(1998)}]{Olshani}%
	\BibitemOpen
	\bibfield  {author} {\bibinfo {author} {\bibfnamefont {M.}~\bibnamefont
			{Olshanii}},\ }\bibfield  {title} {\bibinfo {title} {Atomic {S}cattering in
			the {P}resence of an {E}xternal {C}onfinement and a {G}as of {I}mpenetrable
			{B}osons},\ }\href {https://doi.org/10.1103/PhysRevLett.81.938} {\bibfield
		{journal} {\bibinfo  {journal} {Phys. Rev. Lett.}\ }\textbf {\bibinfo
			{volume} {81}},\ \bibinfo {pages} {938} (\bibinfo {year} {1998})}\BibitemShut
	{NoStop}%
	\bibitem [{\citenamefont {Mistakidis}\ \emph {et~al.}(2023)\citenamefont
		{Mistakidis}, \citenamefont {Volosniev}, \citenamefont {Barfknecht},
		\citenamefont {Fogarty}, \citenamefont {Busch}, \citenamefont {Foerster},
		\citenamefont {Schmelcher},\ and\ \citenamefont {Zinner}}]{MISTAKIDIS20231}%
	\BibitemOpen
	\bibfield  {author} {\bibinfo {author} {\bibfnamefont {S.}~\bibnamefont
			{Mistakidis}}, \bibinfo {author} {\bibfnamefont {A.}~\bibnamefont
			{Volosniev}}, \bibinfo {author} {\bibfnamefont {R.}~\bibnamefont
			{Barfknecht}}, \bibinfo {author} {\bibfnamefont {T.}~\bibnamefont {Fogarty}},
		\bibinfo {author} {\bibfnamefont {T.}~\bibnamefont {Busch}}, \bibinfo
		{author} {\bibfnamefont {A.}~\bibnamefont {Foerster}}, \bibinfo {author}
		{\bibfnamefont {P.}~\bibnamefont {Schmelcher}},\ and\ \bibinfo {author}
		{\bibfnamefont {N.}~\bibnamefont {Zinner}},\ }\bibfield  {title} {\bibinfo
		{title} {Few-body {B}ose gases in low dimensions—a laboratory for quantum
			dynamics},\ }\href
	{https://doi.org/https://doi.org/10.1016/j.physrep.2023.10.004} {\bibfield
		{journal} {\bibinfo  {journal} {Phys. Rep.}\ }\textbf {\bibinfo {volume}
			{1042}},\ \bibinfo {pages} {1} (\bibinfo {year} {2023})}\BibitemShut
	{NoStop}%
	\bibitem [{\citenamefont {Chin}\ \emph {et~al.}(2010)\citenamefont {Chin},
		\citenamefont {Grimm}, \citenamefont {Julienne},\ and\ \citenamefont
		{Tiesinga}}]{Cheng_Feshbach}%
	\BibitemOpen
	\bibfield  {author} {\bibinfo {author} {\bibfnamefont {C.}~\bibnamefont
			{Chin}}, \bibinfo {author} {\bibfnamefont {R.}~\bibnamefont {Grimm}},
		\bibinfo {author} {\bibfnamefont {P.}~\bibnamefont {Julienne}},\ and\
		\bibinfo {author} {\bibfnamefont {E.}~\bibnamefont {Tiesinga}},\ }\bibfield
	{title} {\bibinfo {title} {Feshbach resonances in ultracold gases},\ }\href
	{https://doi.org/10.1103/RevModPhys.82.1225} {\bibfield  {journal} {\bibinfo
			{journal} {Rev. Mod. Phys.}\ }\textbf {\bibinfo {volume} {82}},\ \bibinfo
		{pages} {1225} (\bibinfo {year} {2010})}\BibitemShut {NoStop}%
	\bibitem [{\citenamefont {Zwerger}(2012)}]{ZwergerBook}%
	\BibitemOpen
	\bibinfo {editor} {\bibfnamefont {W.}~\bibnamefont {Zwerger}},\ ed.,\ \href
	{https://doi.org/10.1007/978-3-642-21978-8} {\emph {\bibinfo {title} {The
				{B}CS-{B}EC {C}rossover and the {U}nitary {F}ermi {G}as}}}\ (\bibinfo
	{publisher} {Springer Berlin},\ \bibinfo {address} {Heidelberg, Germany},\
	\bibinfo {year} {2012})\BibitemShut {NoStop}%
	\bibitem [{\citenamefont {Bogoliubov}(1947)}]{bogoliubov1947theory}%
	\BibitemOpen
	\bibfield  {author} {\bibinfo {author} {\bibfnamefont {N.}~\bibnamefont
			{Bogoliubov}},\ }\bibfield  {title} {\bibinfo {title} {On the theory of
			superfluidity},\ }\href@noop {} {\bibfield  {journal} {\bibinfo  {journal}
			{J. Phys.}\ }\textbf {\bibinfo {volume} {11}},\ \bibinfo {pages} {23}
		(\bibinfo {year} {1947})}\BibitemShut {NoStop}%
	\bibitem [{\citenamefont {Ueda}(2010)}]{Ueda-book}%
	\BibitemOpen
	\bibfield  {author} {\bibinfo {author} {\bibfnamefont {M.}~\bibnamefont
			{Ueda}},\ }\href {https://doi.org/10.1142/7216} {\emph {\bibinfo {title}
			{Fundamentals and {N}ew {F}rontiers of {B}ose-{E}instein {C}ondensation}}}\
	(\bibinfo  {publisher} {World Scientific},\ \bibinfo {address} {Singapore},\
	\bibinfo {year} {2010})\ \Eprint
	{https://arxiv.org/abs/https://www.worldscientific.com/doi/pdf/10.1142/7216}
	{https://www.worldscientific.com/doi/pdf/10.1142/7216} \BibitemShut {NoStop}%
	\bibitem [{\citenamefont {Parajuli}\ and\ \citenamefont
		{Chien}(2023)}]{PhysRevA.107.063314}%
	\BibitemOpen
	\bibfield  {author} {\bibinfo {author} {\bibfnamefont {B.}~\bibnamefont
			{Parajuli}}\ and\ \bibinfo {author} {\bibfnamefont {C.-C.}\ \bibnamefont
			{Chien}},\ }\bibfield  {title} {\bibinfo {title} {Proximity effect and
			spatial {K}ibble-{Z}urek mechanism in atomic {F}ermi gases with inhomogeneous
			pairing interactions},\ }\href {https://doi.org/10.1103/PhysRevA.107.063314}
	{\bibfield  {journal} {\bibinfo  {journal} {Phys. Rev. A}\ }\textbf {\bibinfo
			{volume} {107}},\ \bibinfo {pages} {063314} (\bibinfo {year}
		{2023})}\BibitemShut {NoStop}%
	\bibitem [{\citenamefont {Shin}\ \emph {et~al.}(2007)\citenamefont {Shin},
		\citenamefont {Schunck}, \citenamefont {Schirotzek},\ and\ \citenamefont
		{Ketterle}}]{PhysRevLett.99.090403}%
	\BibitemOpen
	\bibfield  {author} {\bibinfo {author} {\bibfnamefont {Y.}~\bibnamefont
			{Shin}}, \bibinfo {author} {\bibfnamefont {C.~H.}\ \bibnamefont {Schunck}},
		\bibinfo {author} {\bibfnamefont {A.}~\bibnamefont {Schirotzek}},\ and\
		\bibinfo {author} {\bibfnamefont {W.}~\bibnamefont {Ketterle}},\ }\bibfield
	{title} {\bibinfo {title} {Tomographic rf spectroscopy of a trapped fermi gas
			at unitarity},\ }\href {https://doi.org/10.1103/PhysRevLett.99.090403}
	{\bibfield  {journal} {\bibinfo  {journal} {Phys. Rev. Lett.}\ }\textbf
		{\bibinfo {volume} {99}},\ \bibinfo {pages} {090403} (\bibinfo {year}
		{2007})}\BibitemShut {NoStop}%
	\bibitem [{\citenamefont {Murthy}\ \emph {et~al.}(2018)\citenamefont {Murthy},
		\citenamefont {Neidig}, \citenamefont {Klemt}, \citenamefont {Bayha},
		\citenamefont {Boettcher}, \citenamefont {Enss}, \citenamefont {Holten},
		\citenamefont {Zürn}, \citenamefont {Preiss},\ and\ \citenamefont
		{Jochim}}]{doi:10.1126/science.aan5950}%
	\BibitemOpen
	\bibfield  {author} {\bibinfo {author} {\bibfnamefont {P.~A.}\ \bibnamefont
			{Murthy}}, \bibinfo {author} {\bibfnamefont {M.}~\bibnamefont {Neidig}},
		\bibinfo {author} {\bibfnamefont {R.}~\bibnamefont {Klemt}}, \bibinfo
		{author} {\bibfnamefont {L.}~\bibnamefont {Bayha}}, \bibinfo {author}
		{\bibfnamefont {I.}~\bibnamefont {Boettcher}}, \bibinfo {author}
		{\bibfnamefont {T.}~\bibnamefont {Enss}}, \bibinfo {author} {\bibfnamefont
			{M.}~\bibnamefont {Holten}}, \bibinfo {author} {\bibfnamefont
			{G.}~\bibnamefont {Zürn}}, \bibinfo {author} {\bibfnamefont {P.~M.}\
			\bibnamefont {Preiss}},\ and\ \bibinfo {author} {\bibfnamefont
			{S.}~\bibnamefont {Jochim}},\ }\bibfield  {title} {\bibinfo {title}
		{High-temperature pairing in a strongly interacting two-dimensional {F}ermi
			gas},\ }\href {https://doi.org/10.1126/science.aan5950} {\bibfield  {journal}
		{\bibinfo  {journal} {Science}\ }\textbf {\bibinfo {volume} {359}},\ \bibinfo
		{pages} {452} (\bibinfo {year} {2018})},\ \Eprint
	{https://arxiv.org/abs/https://www.science.org/doi/pdf/10.1126/science.aan5950}
	{https://www.science.org/doi/pdf/10.1126/science.aan5950} \BibitemShut
	{NoStop}%
	\bibitem [{\citenamefont {Exner}\ \emph {et~al.}(2024)\citenamefont {Exner},
		\citenamefont {Srikumar}, \citenamefont {Blattner}, \citenamefont {Eiles},
		\citenamefont {Schmelcher},\ and\ \citenamefont {Ott}}]{Exner25}%
	\BibitemOpen
	\bibfield  {author} {\bibinfo {author} {\bibfnamefont {M.}~\bibnamefont
			{Exner}}, \bibinfo {author} {\bibfnamefont {R.}~\bibnamefont {Srikumar}},
		\bibinfo {author} {\bibfnamefont {R.}~\bibnamefont {Blattner}}, \bibinfo
		{author} {\bibfnamefont {M.~T.}\ \bibnamefont {Eiles}}, \bibinfo {author}
		{\bibfnamefont {P.}~\bibnamefont {Schmelcher}},\ and\ \bibinfo {author}
		{\bibfnamefont {H.}~\bibnamefont {Ott}},\ }\href@noop {} {\bibinfo {title}
		{High precision spectroscopy of trilobite rydberg molecules}} (\bibinfo
	{year} {2024}),\ \bibinfo {note} {arxiv:2412.19710}\BibitemShut {NoStop}%
	\bibitem [{\citenamefont {Sheludko}\ \emph {et~al.}(2008)\citenamefont
		{Sheludko}, \citenamefont {Bell}, \citenamefont {Anderson}, \citenamefont
		{Hofmann}, \citenamefont {Vredenbregt},\ and\ \citenamefont
		{Scholten}}]{PhysRevA.77.033401}%
	\BibitemOpen
	\bibfield  {author} {\bibinfo {author} {\bibfnamefont {D.~V.}\ \bibnamefont
			{Sheludko}}, \bibinfo {author} {\bibfnamefont {S.~C.}\ \bibnamefont {Bell}},
		\bibinfo {author} {\bibfnamefont {R.}~\bibnamefont {Anderson}}, \bibinfo
		{author} {\bibfnamefont {C.~S.}\ \bibnamefont {Hofmann}}, \bibinfo {author}
		{\bibfnamefont {E.~J.~D.}\ \bibnamefont {Vredenbregt}},\ and\ \bibinfo
		{author} {\bibfnamefont {R.~E.}\ \bibnamefont {Scholten}},\ }\bibfield
	{title} {\bibinfo {title} {State-selective imaging of cold atoms},\ }\href
	{https://doi.org/10.1103/PhysRevA.77.033401} {\bibfield  {journal} {\bibinfo
			{journal} {Phys. Rev. A}\ }\textbf {\bibinfo {volume} {77}},\ \bibinfo
		{pages} {033401} (\bibinfo {year} {2008})}\BibitemShut {NoStop}%
	\bibitem [{\citenamefont {Martinez-Dorantes}\ \emph {et~al.}(2018)\citenamefont
		{Martinez-Dorantes}, \citenamefont {Alt}, \citenamefont {Gallego},
		\citenamefont {Ghosh}, \citenamefont {Ratschbacher},\ and\ \citenamefont
		{Meschede}}]{PhysRevA.97.023410}%
	\BibitemOpen
	\bibfield  {author} {\bibinfo {author} {\bibfnamefont {M.}~\bibnamefont
			{Martinez-Dorantes}}, \bibinfo {author} {\bibfnamefont {W.}~\bibnamefont
			{Alt}}, \bibinfo {author} {\bibfnamefont {J.}~\bibnamefont {Gallego}},
		\bibinfo {author} {\bibfnamefont {S.}~\bibnamefont {Ghosh}}, \bibinfo
		{author} {\bibfnamefont {L.}~\bibnamefont {Ratschbacher}},\ and\ \bibinfo
		{author} {\bibfnamefont {D.}~\bibnamefont {Meschede}},\ }\bibfield  {title}
	{\bibinfo {title} {State-dependent fluorescence of neutral atoms in optical
			potentials},\ }\href {https://doi.org/10.1103/PhysRevA.97.023410} {\bibfield
		{journal} {\bibinfo  {journal} {Phys. Rev. A}\ }\textbf {\bibinfo {volume}
			{97}},\ \bibinfo {pages} {023410} (\bibinfo {year} {2018})}\BibitemShut
	{NoStop}%
	\bibitem [{\citenamefont {Gierling}\ \emph {et~al.}(2011)\citenamefont
		{Gierling}, \citenamefont {Schneeweiss}, \citenamefont {Visanescu},
		\citenamefont {Federsel}, \citenamefont {H{\"a}ffner}, \citenamefont {Kern},
		\citenamefont {Judd}, \citenamefont {G{\"u}nther},\ and\ \citenamefont
		{Fort{\'a}gh}}]{Gierling2011}%
	\BibitemOpen
	\bibfield  {author} {\bibinfo {author} {\bibfnamefont {M.}~\bibnamefont
			{Gierling}}, \bibinfo {author} {\bibfnamefont {P.}~\bibnamefont
			{Schneeweiss}}, \bibinfo {author} {\bibfnamefont {G.}~\bibnamefont
			{Visanescu}}, \bibinfo {author} {\bibfnamefont {P.}~\bibnamefont {Federsel}},
		\bibinfo {author} {\bibfnamefont {M.}~\bibnamefont {H{\"a}ffner}}, \bibinfo
		{author} {\bibfnamefont {D.~P.}\ \bibnamefont {Kern}}, \bibinfo {author}
		{\bibfnamefont {T.~E.}\ \bibnamefont {Judd}}, \bibinfo {author}
		{\bibfnamefont {A.}~\bibnamefont {G{\"u}nther}},\ and\ \bibinfo {author}
		{\bibfnamefont {J.}~\bibnamefont {Fort{\'a}gh}},\ }\bibfield  {title}
	{\bibinfo {title} {Cold-atom scanning probe microscopy},\ }\href
	{https://doi.org/10.1038/nnano.2011.80} {\bibfield  {journal} {\bibinfo
			{journal} {Nature Nanotechnology}\ }\textbf {\bibinfo {volume} {6}},\
		\bibinfo {pages} {446} (\bibinfo {year} {2011})}\BibitemShut {NoStop}%
	\bibitem [{\citenamefont {Kuhr}(2016)}]{10.1093/nsr/nww023}%
	\BibitemOpen
	\bibfield  {author} {\bibinfo {author} {\bibfnamefont {S.}~\bibnamefont
			{Kuhr}},\ }\bibfield  {title} {\bibinfo {title} {Quantum-gas microscopes: a
			new tool for cold-atom quantum simulators},\ }\href
	{https://doi.org/10.1093/nsr/nww023} {\bibfield  {journal} {\bibinfo
			{journal} {National Science Review}\ }\textbf {\bibinfo {volume} {3}},\
		\bibinfo {pages} {170} (\bibinfo {year} {2016})},\ \Eprint
	{https://arxiv.org/abs/https://academic.oup.com/nsr/article-pdf/3/2/170/31565423/nww023.pdf}
	{https://academic.oup.com/nsr/article-pdf/3/2/170/31565423/nww023.pdf}
	\BibitemShut {NoStop}%
	\bibitem [{\citenamefont {Stecker}\ \emph {et~al.}(2017)\citenamefont
		{Stecker}, \citenamefont {Schefzyk}, \citenamefont {Fortágh},\ and\
		\citenamefont {Günther}}]{Stecker_2017}%
	\BibitemOpen
	\bibfield  {author} {\bibinfo {author} {\bibfnamefont {M.}~\bibnamefont
			{Stecker}}, \bibinfo {author} {\bibfnamefont {H.}~\bibnamefont {Schefzyk}},
		\bibinfo {author} {\bibfnamefont {J.}~\bibnamefont {Fortágh}},\ and\
		\bibinfo {author} {\bibfnamefont {A.}~\bibnamefont {Günther}},\ }\bibfield
	{title} {\bibinfo {title} {A high resolution ion microscope for cold atoms},\
	}\href {https://doi.org/10.1088/1367-2630/aa6741} {\bibfield  {journal}
		{\bibinfo  {journal} {New Journal of Physics}\ }\textbf {\bibinfo {volume}
			{19}},\ \bibinfo {pages} {043020} (\bibinfo {year} {2017})}\BibitemShut
	{NoStop}%
	\bibitem [{\citenamefont {Whalen}\ \emph {et~al.}(2019)\citenamefont {Whalen},
		\citenamefont {Kanungo}, \citenamefont {Ding}, \citenamefont {Wagner},
		\citenamefont {Schmidt}, \citenamefont {Sadeghpour}, \citenamefont {Yoshida},
		\citenamefont {Burgd\"orfer}, \citenamefont {Dunning},\ and\ \citenamefont
		{Killian}}]{whalen2019}%
	\BibitemOpen
	\bibfield  {author} {\bibinfo {author} {\bibfnamefont {J.~D.}\ \bibnamefont
			{Whalen}}, \bibinfo {author} {\bibfnamefont {S.~K.}\ \bibnamefont {Kanungo}},
		\bibinfo {author} {\bibfnamefont {R.}~\bibnamefont {Ding}}, \bibinfo {author}
		{\bibfnamefont {M.}~\bibnamefont {Wagner}}, \bibinfo {author} {\bibfnamefont
			{R.}~\bibnamefont {Schmidt}}, \bibinfo {author} {\bibfnamefont {H.~R.}\
			\bibnamefont {Sadeghpour}}, \bibinfo {author} {\bibfnamefont
			{S.}~\bibnamefont {Yoshida}}, \bibinfo {author} {\bibfnamefont
			{J.}~\bibnamefont {Burgd\"orfer}}, \bibinfo {author} {\bibfnamefont {F.~B.}\
			\bibnamefont {Dunning}},\ and\ \bibinfo {author} {\bibfnamefont {T.~C.}\
			\bibnamefont {Killian}},\ }\bibfield  {title} {\bibinfo {title} {Probing
			nonlocal spatial correlations in quantum gases with ultra-long-range rydberg
			molecules},\ }\href {https://doi.org/10.1103/PhysRevA.100.011402} {\bibfield
		{journal} {\bibinfo  {journal} {Phys. Rev. A}\ }\textbf {\bibinfo {volume}
			{100}},\ \bibinfo {pages} {011402} (\bibinfo {year} {2019})}\BibitemShut
	{NoStop}%
	\bibitem [{\citenamefont {Hollerith}\ \emph {et~al.}(2019)\citenamefont
		{Hollerith}, \citenamefont {Zeiher}, \citenamefont {Rui}, \citenamefont
		{Rubio-Abadal}, \citenamefont {Walther}, \citenamefont {Pohl}, \citenamefont
		{Stamper-Kurn}, \citenamefont {Bloch},\ and\ \citenamefont
		{Gross}}]{hollerith2019}%
	\BibitemOpen
	\bibfield  {author} {\bibinfo {author} {\bibfnamefont {S.}~\bibnamefont
			{Hollerith}}, \bibinfo {author} {\bibfnamefont {J.}~\bibnamefont {Zeiher}},
		\bibinfo {author} {\bibfnamefont {J.}~\bibnamefont {Rui}}, \bibinfo {author}
		{\bibfnamefont {A.}~\bibnamefont {Rubio-Abadal}}, \bibinfo {author}
		{\bibfnamefont {V.}~\bibnamefont {Walther}}, \bibinfo {author} {\bibfnamefont
			{T.}~\bibnamefont {Pohl}}, \bibinfo {author} {\bibfnamefont {D.~M.}\
			\bibnamefont {Stamper-Kurn}}, \bibinfo {author} {\bibfnamefont
			{I.}~\bibnamefont {Bloch}},\ and\ \bibinfo {author} {\bibfnamefont
			{C.}~\bibnamefont {Gross}},\ }\bibfield  {title} {\bibinfo {title} {Quantum
			gas microscopy of rydberg macrodimers},\ }\href
	{https://doi.org/10.1126/science.aaw4150} {\bibfield  {journal} {\bibinfo
			{journal} {Science}\ }\textbf {\bibinfo {volume} {364}},\ \bibinfo {pages}
		{664} (\bibinfo {year} {2019})},\ \Eprint
	{https://arxiv.org/abs/https://www.science.org/doi/pdf/10.1126/science.aaw4150}
	{https://www.science.org/doi/pdf/10.1126/science.aaw4150} \BibitemShut
	{NoStop}%
\end{thebibliography}
%

\end{document}